%% file: performance-interaction-detection.tex
\documentclass[acmsmall,nonacm=true]{acmart}

\usepackage{acronym}
\usepackage[linesnumbered,ruled]{algorithm2e}
\usepackage{amstext}
\usepackage{amsthm}
\usepackage{fontawesome}
\usepackage[geometry]{ifsym}
\usepackage{listings}
\usepackage{mathtools}
\usepackage{mdframed}
\usepackage{multicol}
\usepackage{multirow}
\usepackage{siunitx}
\usepackage{subcaption}
\usepackage{tablefootnote}
\usepackage{xfrac}
\usepackage{xspace}

\definecolor{comment}{cmyk}{0.82,0.39,0,0.25}
\usepackage[prependcaption,textsize=footnotesize,linecolor=comment,backgroundcolor=comment!25,bordercolor=comment]{todonotes}

\input{tex/macros}
\begin{document}

\title{Detecting Performance-Relevant Changes in Configurable Software Systems}

\author{Sebastian B\"ohm}
\email{boehmseb@cs.uni-saarland.de}
\orcid{0000-0002-5039-3186}
\affiliation{%
  \institution{Saarland University}
  \city{Saarbr\"ucken}
  \country{Germany}
}

\author{Florian Sattler}
\email{sattlerf@cs.uni-saarland.de}
\orcid{0000-0003-2523-1158}
\affiliation{%
  \institution{Saarland University}
  \city{Saarbr\"ucken}
  \country{Germany}
}

\author{Norbert Siegmund}
\email{norbert.siegmund@uni-leipzig.de}
\orcid{0000-0001-7741-7777}
\affiliation{%
  \institution{Leipzig University}
  \city{Leipzig}
  \country{Germany}
}

\author{Sven Apel}
\email{apel@cs.uni-saarland.de}
\orcid{0000-0003-3687-2233}
\affiliation{%
  \institution{Saarland University}
  \city{Saarbr\"ucken}
  \country{Germany}
}

\renewcommand{\shortauthors}{S.\ B\"ohm, F.\ Sattler, S.\ Apel}

\begin{abstract}
\input{tex/abstract}
\end{abstract}

\maketitle

\input{tex/introduction}
\input{tex/background}
\input{tex/theory}
\input{tex/study}
\input{tex/results}
\input{tex/threats}
\input{tex/related_work}
\input{tex/conclusion}

\bibliographystyle{ACM-Reference-Format}
\bibliography{main}

\end{document}

%% file: tex/macros.tex
\newcommand{\vara}{\textsc{VaRA}\xspace}

\newcommand{\llvmir}{LLVM-IR\xspace}
\newcommand{\approach}{\textsc{ConfFLARE}\xspace}

\definecolor{color-diff}{rgb}{1,.5,.5}
\definecolor{color+diff}{rgb}{.5,1,.5}
\definecolor{change}{rgb}{.5,1,.5}
\definecolor{hot}{rgb}{1,.2,0}
\definecolor{feature1}{rgb}{1,0,1}
\definecolor{feature2}{rgb}{0,0,1}
\definecolor{esghighlight}{HTML}{F58137}
\definecolor{goldweb}{rgb}{1.0, 0.84, 0.0}

\newcommand{\concarrow}{\ensuremath{\raisebox{-2pt}{\text{\FilledDiamondshape}}}\xspace}
\newcommand{\dfrel}{\rightsquigarrow}
\newcommand{\pirel}{\rightsquigarrow_{\text{\tiny\faTachometer}}}
\newcommand{\crset}{C \hspace{-0.06em} R}
\newcommand{\priset}{P \hspace{-0.04em} I}
\newcommand{\changeinst}{i_\text{\faCodeFork}}

\DeclarePairedDelimiter\myset{\{}{\}}

\DeclarePairedDelimiter\abs{\lvert}{\rvert}

\newacro{AST}[AST]{abstract syntax tree}
\newacro{CFG}[CFG]{control-flow graph}
\newacro{IDE}[IDE]{Interprocedural Distributive Environments}
\newacro{IFDS}[IFDS]{Interprocedural Finite Distributive Subset}
\newacro{ESG}[ESG]{exploded super-graph}
\newacro{PGO}[PGO]{profile-guided optimization}
\newacro{SSA}[SSA]{static single assignment}
\newacro{VCS}[VCS]{version control system}
\newacro{bACC}[bACC]{balanced accuracy}

\newcommand{\func}[1]{\ensuremath{#1}}

\theoremstyle{definition}
\newtheorem{definition}{Definition}

\makeatletter
\newcommand{\customlabel}[2]{%
  \@bsphack
  \begingroup
  \def\label@name{#1}%
  \label@hook
  \protected@write\@auxout{}{%
    \string\newlabel{#1}{%
      {#2\textsubscript{\@currentlabel}}%
      {\thepage}%
      {\@currentlabelname}%
      {\@currentHref}{}%
    }%
  }%
  \endgroup
  \@esphack
}
\makeatother

\newcounter{ScenarioCounter}

\newcounter{RQCounter}
\newcommand{\newrq}[1]{%
  \refstepcounter{RQCounter}\par\medskip
  \def\@currentlabel{\arabic{RQCounter}}{}{}%
  \begin{mdframed}[%
    linecolor=black!50,%
    linewidth=3pt,%
    topline=false,%
    bottomline=false,%
    rightline=false,%
    backgroundcolor=black!5,%
    innerleftmargin=8pt,%
  ]
    \begin{tabular}{@{}r@{}l@{}}
      \noindent \textit{\textbf{RQ\textsubscript{\arabic{RQCounter}}:\enspace}}%
      \customlabel{rq:\arabic{RQCounter}}{RQ}%
      & \parbox[t]{0.92\textwidth}{\textit{#1}}
    \end{tabular}
  \end{mdframed}
  \medskip
}

\newcommand{\steprq}{%
  \refstepcounter{RQCounter}\par\bigskip
  \def\@currentlabel{\arabic{RQCounter}}{}{}
}

\newcounter{RQSubCounter}[RQCounter]
\newcommand{\subrq}[1]{%
  \refstepcounter{RQSubCounter}\par\bigskip
  \def\@currentlabel{\arabic{RQCounter}.\arabic{RQSubCounter}}{}{}%
  \noindent \textit{\textbf{RQ\textsubscript{\arabic{RQCounter}.\arabic{RQSubCounter}}:\enspace}}%
  \customlabel{rq:\arabic{RQCounter}.\arabic{RQSubCounter}}{RQ}%
  \textit{#1}\medskip
}

\newcommand{\rqanswer}[2]{%
  \par\medskip
  \begin{mdframed}[%
    linecolor=black!50,%
    linewidth=0.5pt,%
    backgroundcolor=black!5,%
    innerleftmargin=8pt,%
  ]
  \begin{tabular}{@{}rl@{}}
    \textit{\textbf{RQ\textsubscript{#1}}} & \parbox{0.92\textwidth}{#2} \\
  \end{tabular}
  \end{mdframed}
  \medskip
}

\definecolor{basicCC}{HTML}{19177d}
\definecolor{codeComment}{HTML}{408080}
\definecolor{string}{RGB}{255,0,0}
\definecolor{keyword}{HTML}{b00040}

\ifPDFTeX
  \let\savedsfdefault\sfdefault
  \let\savedfamilydefault\familydefault
  \usepackage{FiraSans}
  \let\sfdefault\savedsfdefault
  \let\familydefault\savedfamilydefault
  \newcommand{\codefont}{\fontfamily{FiraSans-OsF}\selectfont}
\else
  \newfontfamily{\codefont}{Fira Sans}
\fi
\newcommand{\code}[1]{\lstinline[style=mycpp, basicstyle=\footnotesize\codefont\color{basicCC}]{#1}}
\newcommand{\captioncode}[1]{\lstinline[style=mycpp]{#1}}

\lstdefinelanguage{cpp}{
  sensitive=true,
  morecomment=[l]//,
  morecomment=[s]{/*}{*/},
  keywords=[1] { bool, char, int, string, unsigned, void, },
  keywords=[2] { return, if, else, for, extern, while, delete, auto, const, static, struct, union, virtual, override },
}

\lstdefinestyle{mycpp} {
  language=cpp,
  basicstyle=\scriptsize\codefont\color{basicCC},
  keywordstyle=[1]\color{keyword},
  keywordstyle=[2]\color{keyword},
  stringstyle=\color{red},
  commentstyle=\color{codeComment},
  numberstyle=\tiny\color{black},
  numbersep=2mm,
  tabsize=8,
  columns=flexible
}

%% file: tex/abstract.tex
Performance is a volatile property of a software system and, as the system evolves, its performance behavior may change, too.
As a consequence, frequent performance profiling is required to keep the knowledge about a software system's performance behavior up to date.
Repeating all performance measurements after every revision is a cost-intensive task, though.
This is even more challenging in the presence of configurability~(e.g., by user-selectable software features), where one has to measure multiple configurations of a given software system to obtain a comprehensive picture.
Configuration sampling is a common approach to control the measurement cost.
However, sampling cannot guarantee completeness and might miss performance regressions, especially if they only affect a small portion of the configuration space.

As an alternative to solve the cost reduction problem, we present \approach:
\approach estimates whether a change potentially impacts performance by identifying data-flow interactions with performance-relevant code.
It then extracts the configurable software features that participate in such interactions, as these features are highly likely to influence how the performance-relevant code is executed.
Based on the identified features, we can select a subset of relevant configurations to focus performance profiling efforts on.

To evaluate the effectiveness of \approach in a performance regression setting, we conduct a study on both synthetic and real-world software systems.
In particular, we have address three questions:
How effective is \approach in detecting performance-relevant changes~(\ref{rq:1}) and in identifying relevant features for performance regression analysis~(\ref{rq:2})?
And how much performance measurement effort can we potentially save with this information~(\ref{rq:3})?
We find that \approach correctly detects performance regressions in almost all cases and identifies the relevant features in all but two cases.
With this information, we are able to reduce the number of configurations to be tested on average by $79\%$ for synthetic and by $70\%$ for real-world regression scenarios saving hours of performance testing time.

%% file: tex/introduction.tex
\section{Introduction}
\label{sec:introduction}

The performance behavior of a software system is crucial for its success.
The challenge is that it is a volatile property that changes frequently as the system in question evolves.
Software evolution can easily lead to performance regressions that negatively affect user experience or operating costs~\cite{JB12}.

Performance regression testing aims at revealing performance issues early such that they can be fixed before deployment~\cite{SSE+25}.
Since running performance tests for every revision is often a cost-intensive task, it is desirable to estimate whether and how incoming changes affect the system’s performance behavior.
Such an estimation prior to actual testing becomes even more important in the presence of configurability:
If a software system is configurable---and most systems are configurable---it is not sufficient to test only multiple revisions, but also multiple configurations need to be tested as well to obtain a complete picture.
This causes an exponential blowup of the number of performance measurement runs.
As a result, continuous performance testing on every revision is infeasible for many configurable software systems such that only a sample of all configurations are tested instead~\cite{MAS+20}.
The problem is that the success of configuration sampling relies on the chance that regressing configurations are contained in the selected sample.
As a consequence, regressions can easily be missed, especially if they affect only a small fraction of the overall configuration space~\cite{MAS+20}.
Many sampling strategies have been devised that offer different trade-offs between sample size and coverage~\cite{KGSA20, MKR+16}.
These strategies typically only consider properties of the configuration space, that is, which configurable software features can be combined in which way, but they do not take advantage of the regression setting because they lack knowledge about which configurations are affected by a change.

Studying performance bugs in highly configurable software systems, \citet{HY+16} suggest to use static analysis to determine which configurations are impacted by performance-relevant code changes so that performance testing efforts can be focused on impacted configurations.
In this work, we present \approach, an approach that achieves exactly that:
In a first step, we employ a data-flow analysis to determine whether a change to the code base of a configurable software system interacts with performance-relevant code in the system.
Only if such an interaction exists, we consider the change for performance analysis.
In that case, we localize in a second step the part of the configuration space that is potentially affected by that change so that fewer configurations need to be analyzed.
As with sampling, \approach is a heuristic approach that trades completeness for scalability (a change may affect performance-relevant code even without data flow).
However, it is a heuristic that is informed by the program's semantics and, thus, can more precisely select relevant configurations than traditional configuration sampling strategies.

\begin{figure}
  \hrule
  \includegraphics{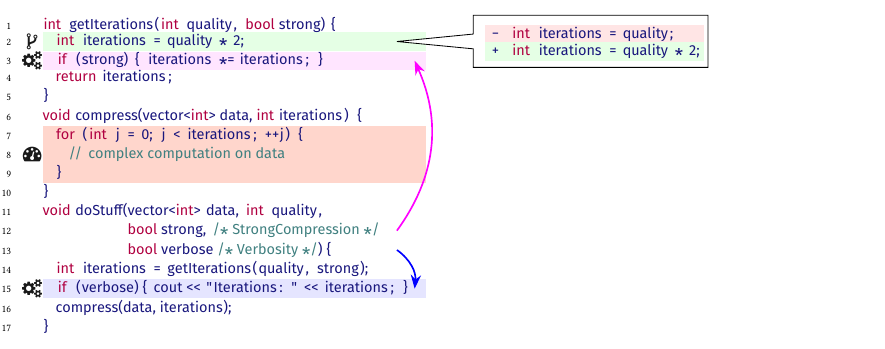}
  \hrule
  \caption{%
    Example of a performance-relevant code change.
    Function \captioncode{compress} contains performance-relevant code (\faTachometer).
    The callout points to a change that was made to Line 2 (\,\faCodeFork\,).
    This change has an influence on the performance behavior of region~\faTachometer\, since it changes how the loop counter is calculated.
    The example contains two features~(\faGears) named \emph{StrongCompression} and \emph{Verbosity}, controlled by the parameters \captioncode{strong} and \captioncode{verbosity}, respectively.
    \approach reveals that only \emph{StrongCompression} affects the example's performance by influencing the value of \captioncode{iterations}.
  }
  \label{fig:intro_example}
\end{figure}

For illustration, consider a typical example of a code change that introduces a performance regression in \autoref{fig:intro_example}, which we use throughout the paper to explain how \approach works.
Here, we briefly outline our idea on solving this problem.
First, it is important to note that we focus here on performance regressions that are caused due to a change in a program's execution.
We do not consider performance regressions that are caused by other external factors or side effects (e.g., I/O, cache effects, etc.).
The example models a simple configurable compression tool that undergoes a change.
It consists of a function \code{doStuff} that takes a vector of data elements as well as some additional parameters and compresses the data vector in two steps:
First, \code{getIterations} computes the number of \code{iterations} that is later used by the compression algorithm.
Second, the data get compressed by function \code{compress}.
This function mainly consists of a loop that is executed for the previously calculated number of \code{iterations}.
As the program evolves, a new revision changes how variable \code{iterations} is computed~(\,\faCodeFork\,).
To judge whether this could lead to a performance regression, we need to decide whether the changes of the new revision influence performance-relevant code, that is code that is central to the program's performance behavior.
\approach does so by determining whether a change interacts with performance-relevant code via a flow of data from the changed code to performance-relevant code.
We call such data-flow interactions \emph{performance-relevant interactions}.
If we do not detect such interactions, we conclude~(heuristically) that the change is likely not performance-relevant and, thus, no performance tests are run.
By contrast, if \approach does find a performance-relevant change, it localizes the part of the configuration space that is potentially affected by that change.
It does so by determining which configurable software features are involved in performance-relevant interactions.
The idea is that only features that affect data flowing from changed code to performance-relevant code can also affect its performance behavior.
Given that a configuration can be characterized by which features are enabled or disabled, configurations involving such features are of special interest for performance testing.
In the example, the configuration space is made up by two features~(\faGears) \emph{StrongCompression} and \emph{Verbosity}.
Only \emph{StrongCompression} affects performance-relevant code by influencing the value of \code{iterations}.
Feature \emph{Verbosity} only reads data and, thus, has no influence on performance.
As a result, we only test one configuration for each possible value of \emph{StrongCompression}, reducing the number of tests to be executed~(from $4$ to $2$).

To evaluate the effectiveness of \approach in a performance regression scenario, we have conducted an empirical study using both synthetic and real-world regression scenarios.
In particular, we have addressed three questions:
How effective is \approach in detecting performance-relevant changes~(\ref{rq:1}) and in identifying relevant features for performance regression analysis~(\ref{rq:2})?
How much performance measurement effort can we potentially save with this information~(\ref{rq:3})?
We found that \approach correctly detects performance regressions in almost all cases and identifies the relevant features in all but two cases.
With this information, we are able to reduce the number of configurations to be tested on average by $79\%$ for synthetic and by $70\%$ for real-world regression scenarios.
In most cases, this leads to speedups of $2$ to $32$ in time spent for performance measurements compared to measuring all configurations, sometimes saving hours of performance testing time.

In summary, we make the following contributions:
\begin{itemize}
  \item We present \approach, an approach that determines whether a change to a configurable software system interacts with performance-relevant code and---if so---computes which part of the configuration-space is affected.
  \item We provide an implementation of our approach in the \vara framework~\cite{SBS+23}.
  \item We evaluate \approach using synthetic and real-world software systems investigating its applicability and effectiveness.
\end{itemize}

All results and a replication package are available online~\footnote{https://anonymous.4open.science/r/ConfFLARE-Supplements}.

%% file: tex/background.tex
\section{Background}
\label{sec:background}

Before we present \approach in detail in \autoref{sec:approach}, we introduce fundamental concepts from previous research.

\subsection{Code Regions}
\label{sub:code_regions}

Code regions are a uniform abstraction to facilitate the common analysis of different kinds of variability~\cite{S23}.
The idea of code regions is to attach high-level domain-specific information to individual program elements, so that results provided by a low-level analysis can later be contextualized in a high-level, domain-specific manner.
For example, we can attach to a piece of code the information by which commit it has been introduced.
A program analysis can then compute the data flows between these code regions to identify commits that interact with each other~\cite{S23, SBS+23}.

Notably, the notion of code regions allows us to represent the three different concepts relevant to \approach---change, performance-relevant code, and features---in a uniform way.
As a result, we can employ existing code-region-based analyses, as we illustrate in \autoref{sub:interaction_analysis}.
Specifically, we use two existing instances of code regions:
(1)~a \emph{commit region} comprises code introduced by a commit and (2)~ a \emph{feature region} denotes code that belongs to a certain feature.
Such a feature region is usually guarded by configuration switches (e.g., feature flags or toggles).
Additionally, we introduce a new kind of code region to model performance-relevant code.

In what follows, we give a formal introduction of code regions, as defined by \citet{S23}, and introduce commit and feature regions as relevant instances.
We model a program $p$ as a set of functions $f_1, \dots, f_n \in p$, where each function is represented as a \ac{CFG}.
We use an instruction-based representation of functions instead of a basic-block-based representation because it allows for a more fine-grained definition of code regions.
Each function $f$ is represented as a triple $\left(\func{entry}_f, \func{exit}_f, \func{insts}(f)\right)$, where $\func{insts}(f)$ is the set of instructions of $f$ and $\func{entry}_f, \func{exit}_f \in \func{insts}(f)$ are the function's unique entry and exit instructions.
The ordering of instructions inside a function is defined by two functions $\func{pred}$ and $\func{next}$.
In our implementation~(see \autoref{sec:approach}), we use \llvmir as a program representation, such that $\func{pred}$ and $\func{next}$ are specified by the semantics of \llvmir.

Since we investigate the evolution of programs, we need to refer to a program revision at a certain point in time.
We use a revision specifier $p^v$ to refer to a program $p$ at revision $v$.
Note that we omit the revision specifier if it is irrelevant in a given context.

To extract code regions from this representation, a tagging function $\func{tags} : I \rightarrow \mathcal{P}(T)$ associates each instruction in the program's set of instructions~$I$ with a set of tags from~$T$.
The tags represent the higher-level concepts (e.g., commit identifiers) that we want to attach to an instruction.
A code region $cr$ is then a connected subgraph of some function $f \in p$ such that all instructions in the code region are tagged with the same tags.
Here, \emph{connected} means that, for any instruction $i$ that belongs to $cr$, there exists another instruction $i' \neq i$ in $cr$ such that $i' \in \func{next}(i) \lor i' \in \func{prev}(i)$.
In other words, a code region is a \emph{contiguous} set of instructions of a function that belong to the same high-level domain-specific concept labelled by a tag.
The set $\crset$ contains all code regions.
Given a code region $cr \in \crset$, we can retrieve the tags associated with that code region using function $\func{regionTags}(cr)$ and the set of instructions using $\func{insts}(cr)$.
For brevity, we use the notation $i \in cr$ instead of $i \in \func{insts}(cr)$.

\paragraph{Commit Regions}

Commit regions comprise instructions that have been introduced by a certain revision in a version control system~\citet{SBS+23}.
The tagging function for commit regions annotates each instruction with the commit that last modified the source code that the instruction is generated from (e.g., using git-blame\footnote{Git-blame is a \ac{VCS} mechanism that annotates each line in a file with the commit that last modified it.}).
We denote the set of all commit regions in a program with $\crset_C$.
We use commit regions to determine which code belongs to a (potentially performance-relevant) change.

\paragraph{Feature Regions}

To model which instructions belong to a certain feature, we use feature regions~\cite{S23}.
We denote the set of all feature regions with $\crset_F$.
The tagging function for feature regions associates each instruction with the name(s) of the feature(s) that specify(ies) whether this instruction is executed.
That is, an instruction $i$ is tagged as belonging to feature $f$ if whether $i$ is executed or not depends on the selection or deselection of $f$.
There are different approaches on how to obtain this information depending on the mechanism that is used for implementing features.
For example, there is a variety of approaches to handle configurability expressed with the \textsc{C} preprocessor~\cite{RTS+16,LJG+15,GG12,KGR+11,LAL+10}  and C++ templates~\cite{GKE+21}.
For load-time configuration options, a static taint analysis can be used to track feature choices that control decisions in the program~\cite{VJS+20,RTS+16,LKB14}. 
We focus on load-time configuration options since this is a very common and practically relevant means to express configurability~\cite{ABK+13}.

\subsection{Region-Based Interaction Analysis}
\label{sub:interaction_analysis}

As outlined in \autoref{sec:introduction}, we aim at detecting performance-relevant changes by identifying data-flow interactions between changes and performance-relevant code.
That is, as we model code changes and performance-relevant code as code regions, \approach requires an analysis that is able to identify data-flow interactions between different code regions.
For this purpose, we use \textsc{SEAL}~\cite{SBS+23}, which uses a state-of-the-art data-flow analysis to compute data-flow interactions between all instructions in a program.
Based on the commit tags attached to the instructions, it infers code from which commits interacts with each other via data flow.

This kind of interaction analysis can be generalized to work with arbitrary code regions.
The idea is that analyses should be decoupled from the domain-specific information associated with concrete code regions, this way, making the analysis reusable for different analysis scenarios.

An interaction between two code regions is based on a given relation $\concarrow \subseteq \crset \times \crset$.
A code region $r_1 \in \crset$ interacts with another code region $r_2 \in \crset$ if and only if $r_1 \neq r_2$ and $r_1 \concarrow r_2$.
The abstract relation $\concarrow$ links region information with the analysis providing a domain-specific view on the general analysis semantics.
To apply this interaction definition to data-flow interactions, we instantiate the abstract relation $\concarrow$ with a concrete data-flow relation $\dfrel$.
That is, $r_1 \dfrel r_2$ iff data flows from some instruction in $r_1$ to some instruction in $r_2$.

Since the concept of region interaction analysis is designed to be independent of the concrete kinds of code regions and only requires appropriate tagging functions, we can use the existing analysis from \textsc{SEAL} with some adaptions to trace data-flow interactions between a change and performance-relevant code.
In the remaining section, we explain our instantiation of this analysis in more detail.

\subsection{Dataflow Analysis and Exploded Supergraph}
\label{sub:esg}

The analysis described in \autoref{sub:interaction_analysis} uses an existing taint analysis based on \acfi{IDE}~\cite{SRH96}, an algorithmic framework for implementing data-flow analyses.
For more details on the concrete analysis implementation, we refer the reader to~\citet{SBS+23}.
One important aspect of \ac{IDE} is the main data structure used to solve data-flow problems---the \emph{\ac{ESG}}.
Since we use the \ac{ESG} later on to find features related to a performance-relevant change, we provide a brief introduction on how an ESG is constructed.

An \ac{ESG} is a graph representation of a program that is used to efficiently solve data-flow problems such as our interaction analysis~\cite{SRH96}.
The nodes in an \ac{ESG} can be written as tuples $(i, d)$, where $i$ is an instruction and $d$ is a data-flow fact.
In our case, $d$ can be a program variable or an instruction that uses another variable.
We consider instructions as data-flow facts to correctly model the static single assignment form of \llvmir.\footnote{In \emph{static single assignment form}, every variable is assigned exactly once. To fulfill this constraint, \llvmir generates temporary variables that are represented by the instructions they stem from. Note that these temporary variables are also data-flow facts.}
\autoref{fig:esg} shows a simplified \ac{ESG} of the program from our example in \autoref{fig:intro_example} next to the corresponding \ac{CFG}s.
For each CFG node (each represents roughly one instruction), we show the most important \ac{ESG} nodes next to it.
Each column stands for a different data-flow fact.
The edges between ESG nodes visualize how data-flow facts propagate throughout the program.
For example, the edge from $(\text{1}, \text{quality})$ to $(\text{2}, \text{iterations})$ reflects how $\text{quality}$ is used to calculate $\text{iterations}$ in Line 2.

%% file: tex/theory.tex
\section{Introducing \approach}
\label{sec:approach}

In this section, we introduce and explain \approach, our approach to detect performance-relevant changes and the corresponding affected part of the configuration-space.
First, we give an overview of \approach using our running example.
Then, we define what we consider performance-relevant code and how we model it as a specific type of code region.
Finally, we describe how \approach computes performance-relevant interactions and how we pin down features that participate in a performance-relevant interaction.

\subsection{Overview}
\label{sub:approach_overview}

\begin{figure}
  \hrule
  \includegraphics[width=0.9\textwidth]{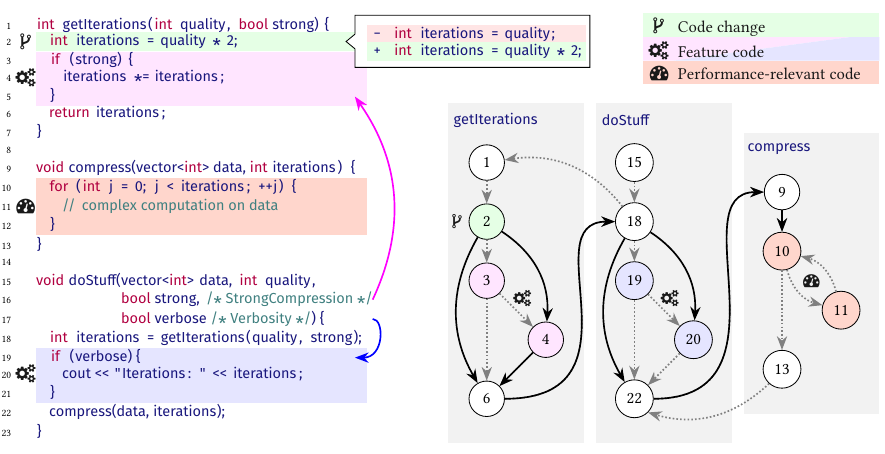}
  \hrule
  \caption{%
    Continuation of our running example from \autoref{fig:intro_example}.
    The code on the left is identical to \autoref{fig:intro_example}.
    On the right is the corresponding control-flow graph with line numbers as nodes and dashed arrows showing control-flow.
    The thick black arrows denote data flow between the change and performance-relevant code passing through \emph{StrongCompression} and \emph{Verbosity}.
  }
  \label{fig:full_example}
\end{figure}

To motivate and explain the idea behind \approach, we continue our running example in \autoref{fig:full_example}.

\paragraph{Performance-Relevant Interactions}
Let's assume that the loop in function \code{compress} dominates the performance behavior of this system~(\faTachometer).
We call this \emph{performance-relevant code} because any change to its performance behavior affects the whole program's performance.
Consequently, if changes from a revision \emph{interact} with such code, then this revision potentially affects performance.
There are many possibilities of how such an interaction could look like.
For example, an interaction could be structural (e.g., the revision directly changes performance-relevant code).
Or an interaction could arise from a function call (e.g., performance-relevant code calls code that has been changed in a revision).
These are rather simple examples that are easy to pin down.
We focus on a more complex kind of interaction, \emph{data-flow interactions}, that occur if data flows from changed code to performance-relevant code.
Data-flow interactions are interesting for two reasons:
(1)~they can be indirect, meaning that there is no obvious relationship between changed code and performance-relevant code~\cite{SBS+23} and
(2)~they are hard to detect with state-of-the art approaches, because this requires a sophisticated data-flow analysis in combination with domain-specific information about program revisions as well as performance-relevant code, which is exactly what we provide with \approach.

The change in our example interacts with performance-relevant code via data flow by modifying how variable \code{iterations} is calculated.
This variable controls the number of iterations the loop in the performance-relevant code region executes.
In \autoref{fig:full_example}, the data-flow paths causing this interaction are denoted by thick black arrows.
The existence of this data-flow interaction between the change and the performance-relevant code indicates that the performance behavior of the system may be affected by that change.
We call such a data-flow interaction a \emph{performance-relevant interaction} and the involved change a \emph{performance-relevant change}.
We use this notion of a performance-relevant interaction to judge whether a given change potentially influences the performance behavior of a system.
In our example, the change causes the loop in \code{compress} to be executed twice as much as previously.
So it can be expected that it takes about twice as long to run the program (assuming that the time spent in the rest of the code is negligible).

\paragraph{Restricting the Configuration Space}
A system's performance behavior often depends on its configuration~\cite{KMG+23}.
Therefore, the entire configuration space must be measured to obtain a comprehensive picture of the performance behavior.
However, this can be very expensive or even intractable for practical systems that provide thousands of configuration options~\cite{ABK+13}.
Sampling is a common approach to control this cost~\cite{KGSA20, MKR+16}.
The problem is that, if a performance regression only affects a small portion of the configuration space, it might be easily missed by sampling~\cite{KMG+23,MAS+20}.
If we are able to identify which part of the configuration space is affected by a performance-relevant change, we can focus measurement efforts only on the affected part, reducing overall measurement costs.

\approach can identify which part of the configuration space is affected by a performance-relevant change by computing which features are \emph{related} to a performance-relevant change.
We deem a feature related if its code participates in data flow from a performance-relevant change to a performance-relevant code region.
Note that \approach does not tell us whether a performance effect occurs when selecting or de-selecting the feature, or for both conditions.
Computing the exact conditions (selections and de-selections of features) under which an effect occurs requires more expensive techniques such as symbolic execution.
Consequently, \approach cannot give any guarantees regarding completeness---similar to sampling.
However, by reducing the configuration space to a part that we know is related to a performance-relevant interaction, \approach can be used to increases the chance that a performance regression is detected, for example by guiding a sampling procedure to focus on this part of the configuration space.

In our example, if we follow the data-flow paths of the interaction, they pass through a node that belongs to the feature \emph{StrongCompression} (i.e., Line 4).
We can also see data flowing from the change to a node related to the feature \emph{Verbosity}~(Line 20).
However, these data are not passed further on to performance-relevant code.
Therefore, we conclude that feature \emph{StrongCompression} is potentially affected by the performance-relevant change, but \emph{Verbosity} is not.

\subsection{Performance-Relevant Code}
\label{sub:perf_region}

There are various reasons for why the performance behavior of a certain piece of code might be important to a stakeholder.
For example, one might want to ensure that the implementation of a computation-heavy algorithm does not regress over time.
In security applications, it is often desirable that certain algorithms take a fixed amount of time to mitigate timing attacks~\cite{ABB+16}.
In safety-critical systems, parts of the code may have to fulfill real-time requirements that could be violated by a change~\cite{MP09}.
In other words, what constitutes performance-relevant code is application-specific such that the user has to define what code they want to consider.
So, we do not restrict \approach to only one kind of performance-relevant code.
Naturally, one might want to automate this declaration for common scenarios.
So, for our evaluation, we chose hot code---code that contributes disproportionately to the overall runtime compared to its size---as performance-relevant code.
This is reasonable in a performance regression setting in which we want to detect performance change-points that arise from changes introduced by new program revisions~\cite{MAS+19}.

In \approach, we model performance-relevant code with its own type of code region:

\begin{definition}
  A \emph{performance region} $cr_P \in \crset_P$ is a code region that contains only instructions belonging to performance-relevant code.
  The tagging function for performance regions tags all instructions that belong to the class of performance-relevant code we are interested in, no additional data are attached:

  \smallskip
  \begin{tabular}{lccl}
    $\func{tags}_P$    & $:$ & \multicolumn{2}{l}{$I \rightarrow \mathcal{P}(\mathbb{B})$}\\
    $\func{tags}_P(i)$ & $=$ & $\{true\}$  & if $i$ is performance-relevant code\\
                       &     & $\emptyset$ & Otherwise\\
  \end{tabular}
\end{definition}

\subsection{Performance-Relevant Interactions}
\label{sub:perf_inter}

We identify potentially performance-relevant interactions by computing how data flow between commit regions associated with changed code and performance regions.
For this purpose, \approach builds on the region-based interaction analysis framework presented in \autoref{sec:background}.
This way, we are able to connect the high-level performance regression scenario with the low-level concept of data flow via the concept of code regions.
The changes made to the program by new revisions map to commit regions, whereas performance-relevant code is represented by performance regions.

To determine which commit regions belong to a certain change, we introduce a projection of the set of all commit regions $\crset_C$ to include only code regions tagged with certain revisions.

\begin{definition}
  Given a set of revisions $V$, the set $\crset_C^{V} \subseteq \crset_C$ contains exactly the commit regions that are tagged with a revision from $V$:
  \begin{flalign*}
    \crset_C^{\textsf{V}} = \big\{ cr\: \big|\: cr \in \crset_C\, \land\, \func{regionTags}(cr) \cap V \neq \emptyset \big\}
  \end{flalign*}
\end{definition}

For a regression scenario in which every new program revision~$v$ is analyzed (e.g., in a CI/CD pipeline), one would choose $V = \{v\}$.

\approach allows us to use the region-based interaction analysis presented in \autoref{sub:interaction_analysis} with multiple different kinds of code regions by slightly adapting the interaction relation $\dfrel$.
Our version of $\dfrel$ restricts the relation such that it only captures data flows from the commit regions in question to performance regions:

\begin{definition}
  Given a set of revisions $V$, the relation $\pirel\; \subseteq \crset_C^V \times \crset_P$ represents data-flow interactions between commit regions that are tagged with revisions from $V$ and performance regions.
\end{definition}

This relation captures only interactions between changes from the desired revisions and performance-relevant code.
Based on this, we can define set of performance-relevant interactions.

\begin{definition}\label{def:performance_relevant_interaction}
  A \emph{performance-relevant interaction} is a tuple consisting of a performance region $cr_P \in \crset_P$ and a set of commit regions that interact with $cr_p$. $\priset$ denotes the set of all performance-relevant interactions.
  \begin{flalign*}
    \priset = \Big\{ \big(cr_P,\ &\func{interactCR}(cr_P) \big)\:\,\Big|\:\,cr_P \in \crset_P\: \land\: \big|\,\func{interactCR}(cr_P)\,\big| > 0 \Big\}
  \end{flalign*}
  where
  \begin{flalign*}
    \func{interactCR}(cr_P) = \big\{ cr_C\:\,\big|\:\,cr_C \in \crset_C^V\: \land\: cr_C \pirel cr_P \big\}
  \end{flalign*}
\end{definition}

A performance-relevant interaction defines which exact parts of a change interact with a certain performance region via data flow.
With this representation, it is easy to see which changes are performance relevant (i.e., interact with performance-relevant code), and it provides the necessary information on which features are related to the interaction, which we describe next.

\subsection{Extracting Feature Information}
\label{sub:feature_info}

Based on the performance-relevant interactions detected in the previous step, we identify a subset of features of a software system that are related to a performance-relevant interaction.
These features are potentially affected by a performance-relevant change and, thus, should be considered when selecting configurations during in-depth performance analysis.
With this goal in mind, we define the criteria that a feature must fulfill to be considered related to a performance-relevant interaction.

\begin{definition}
  \label{def:relevant_feature}
  A feature $f \in F$ is \emph{related to a performance-relevant change} if, at least, one instruction $i$ of a feature region $cr_f$ associated with $f$ fulfills the following criteria:
  \begin{flalign*}
    \text{Let}\;\; cr_C \in \crset_C^V,\: cr_P \in \crset_P,\: cr_C \pirel cr_P. &&
  \end{flalign*}
  \begin{enumerate}
    \item $i$ is part of a program path $\pi$ that gives rise to a performance-relevant interaction:
          \begin{flalign*}
            i \in \pi,\quad \pi = i_1, i_2, \cdots, i_n,\quad i_1 \in cr_C,\: i_n \in cr_P
          \end{flalign*}
    \item $i$ uses data that flow from the change to the performance-relevant code region:
          \begin{flalign*}
            \exists\ \changeinst.\:\: \changeinst \dfrel i\,\land\,\changeinst \in cr_C
          \end{flalign*}
  \end{enumerate}
\end{definition}

We denote the set of all features that fall under this definition with~$\hat{F}$.
To identify all features~$f \in \hat{F}$, we extract the feature names from all feature regions that contain, at least, one instruction that fulfills both criteria.

Checking criterion~(1), we perform a backwards search in the \ac{ESG} constructed by the interaction analysis (see.~\autoref{sub:esg}).
We formalize this procedure in \autoref{alg:feature_identification}.
To collect all instructions that are part of a program path that gives rise to a performance-relevant interaction, we start the traversal at nodes $(i, i)$ belonging to instructions $i \in cr_P$ (Line~2).
These nodes represent performance-relevant code that is affected by a change.
In our example, $(\text{10c}, \text{10c})$ is the only such node.
For each start node, \FuncSty{reconstructPath} progresses backwards along the \ac{ESG}'s edges until it reaches an edge that leads out of a commit region that is part of the performance-relevant change under investigation (Lines~3,~16).
In our example, this happens when we reach the node $(\text{2}, \text{iterations})$.
In addition, we prune all sub-paths where the search terminates without ever encountering a node that is part of such a commit region (Line~11).
This process, which is an instantiation of program slicing~\cite{W84}, collects the instructions that fulfill criterion~(1).

\begin{algorithm}
  \small
  \DontPrintSemicolon
  \SetAlgoVlined

  \SetKwProg{Fn}{fun}{}{end}
  \SetKw{In}{in}
  \SetKwFunction{Just}{Just}
  \SetKwFunction{Nothing}{Nothing}
  \SetKw{Continue}{continue}
  \SetKwFunction{extractFeatures}{reconstructPaths}
  \SetKwFunction{reconstructPath}{reconstructPath}
  \SetKwFunction{predecessors}{pred}
  \SetKwFunction{getFeatures}{getFeatures}

  \SetKwData{Features}{features}
  \SetKwData{Node}{node}
  \SetKwData{Path}{path}
  \SetKwData{Result}{result}
  \SetKwData{Pred}{pred}
  \SetKwData{PredInsts}{predInsts}
  \SetKwData{PrunePath}{prunePath}
  \SetKwArray{Inst}{inst}
  \SetKwArray{Insts}{insts}
  \SetKwArray{StartNodes}{$startNodes$}
  \SetKwArray{EndNodes}{endNodes}

  \Fn{\extractFeatures{$cr_P$}}{
    \StartNodes $\coloneqq \left\{ (i, i)\: \big|\: i \in cr_P \right\}$\;
    \EndNodes $\coloneqq \left\{ (i, d)\: \big|\: \exists\ cr_C.\:\: cr_C \in \func{interactCR(cr_P)} \land i \in cr_C \right\}$\;
    \Insts $\coloneqq \emptyset$\;

    \For{\Node $\in$ \StartNodes}{
      \Result $\coloneqq$ \reconstructPath{\Node, \EndNodes}\;

      \If{\Result $= \left(\Just~\Path\right)$}{
        \Insts $\coloneqq$ \Insts $\cup$ \Path\;
      }
    }

    \Return{\Insts}\;
  }
  \BlankLine
  \Fn{\reconstructPath{\Node, \EndNodes}}{
    \If{$\neg (\Node \in \EndNodes) \land \predecessors{\Node} = \emptyset$}{
      \Return{\Nothing}\;
    }

    \PrunePath $\coloneqq$ true\;
    \Insts $\coloneqq$ $\left\{\Node\right\}$\;

    \For{\Pred $\in$ \predecessors{\Node}}{
      \If{$\Node \in \EndNodes \land \neg\Pred \in \EndNodes$}{
        \Continue\;
      }

      \Result $\coloneqq$ \reconstructPath{\Pred, \EndNodes}\;

      \If{\Result $= \left(\Just~\PredInsts\right)$}{
        \PrunePath $\coloneqq$ false\;
        \Insts $\coloneqq$ \Insts $\cup$ \PredInsts\;
      }
    }

    \If{\PrunePath}{
      \Return{\Nothing}
    }
    \Return{\Just \Insts}\;
  }

  \caption{Reconstruct data-flow paths for a performance-relevant interaction.}
  \label{alg:feature_identification}
\end{algorithm}

\begin{figure}
  \hrule
  \vspace{-1.2ex}
  \includegraphics{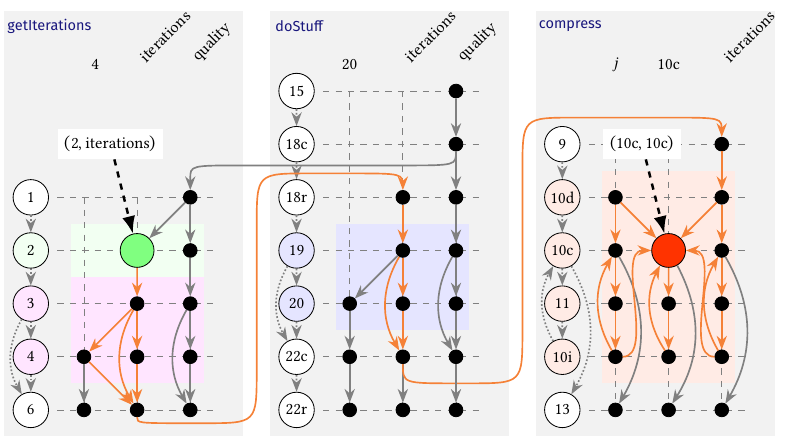}
  \hrule
  \caption{%
    Simplified \ac{ESG} for our running example.
    For each function, we show its \ac{CFG} with line numbers as nodes on the left and a simplified version of its \ac{ESG} on the right that only contains nodes relevant for this example.
    Function calls are split into separate call and return nodes (suffixes \emph{c} and \emph{r}).
    The loop header in Line 10 is split into three nodes: variable declaration (10d), loop condition (10c), and increment (10i).
    The highlighted arrows show the result of the path reconstruction for the performance-relevant interaction between the change in Line 2 and the performance-relevant code region in Lines 10 and 11.
  }
  \label{fig:esg}
\end{figure}

Criterion~(2) further restricts the set of instructions obtained in the previous step to include only those instructions that \emph{use} data that originate from the change and eventually arrive at the performance-relevant code region.
Note that, without any additional filter, the result from \autoref{alg:feature_identification} still includes instructions that load data, but neither modify nor use them for control-flow decisions.
These instructions might bring additional features into consideration.
This can be useful in certain scenarios, for example, in security applications where side-channel attacks are a concern.
However, for our use case of performance regression analysis, such read-only instructions are not expected to play a considerable role in a program's performance behavior.
So, we opt for keeping the configuration space as small as possible by excluding them.

We can observe the effects of applying the additional filters in our example:
By applying criterion (1), we receive the following set of instructions:
\[\myset*{\text{2}, \text{3}, \text{4}, \text{6}, \text{9}, \text{10d}, \text{10c}, \text{10i}, \text{11}, \text{18r}, \text{19}, \text{20}, \text{22c}}\]
From this set, only instructions 4 and 20 are associated with a feature, and they both interact with instruction 2, meaning that we would report the features \emph{StrongCompression} and \emph{Verbosity} as related to the performance-relevant interaction.
By excluding the read-only interaction of instruction 20, we can omit the feature \emph{Verbosity}, since it has no influence on the value that is propagated to the performance-relevant region.
With this extra filter, \approach can be fine-tuned with regard to how aggressively it should restrict the configuration space depending on the use case.

Finally, we can use $\hat{F}$ to select a subset $\hat{C}$ from the set of configurations of a software system $C$.
One possible way of creating such a subset is to cover all possible combinations of the features in $\hat{F}$.
This is a very conservative approach to creating $\hat{C}$ that considers many configurations, but serves as a good reference point for discussing how much the information provided by \approach helps reduce performance testing costs.

Note that, in practice, other means of creating $\hat{C}$ can be chosen.
For example, $\hat{F}$ could be used to guide a sampling strategy to explore a certain part of the configuration space.

%% file: tex/study.tex
\section{Empirical Study}
\label{sec:study}

To evaluate the applicability and capabilities of \approach in a performance regression setting, we conduct a study using both synthetic and real-world software systems.
In particular, we investigate how effective our approach is in detecting performance-relevant changes and in identifying relevant features for performance regression analysis and how much this information helps reduce the cost of such analysis.

\subsection{Research Questions}
\label{sub:research_questions}

Our goal with \approach is to reduce the costs of performance regression analysis of configurable software systems by identifying changes and configurations that potentially cause performance regressions.
To evaluate to which extent we achieve this goal, we address three research questions.
First, we look into whether our approach selects appropriate changes for performance testing:

\newrq{How well does \approach identify changes that incur a performance regressions?}

Here, we focus on the first step of \approach, which identifies whether a change is potentially performance-relevant.
To be effective, it should find performance-relevant interactions for all changes that cause a performance regression.

\medskip
Second, we investigate whether \approach provides useful information for selecting appropriate configurations for performance testing:

\newrq{How accurately does \approach identify features related to performance regressions?}

This question focuses on the second step of \approach, which identifies features related to a performance-relevant change.
Correctly identifying such features is important as any missed feature might cause configurations that are necessary to detect a performance regression to be omitted from performance tests.
To check whether all relevant features were indeed identified, we select a subset of configurations based on the relevant features and check whether we can still detect the performance regression using only this set of configurations.

\medskip
Finally, we take a look at the outcome of \approach in terms of cost savings for performance regression analysis:

\newrq{How much performance measurement effort can be potentially saved with the information provided by \approach?}

Our overall goal is to reduce the costs of performance regression analysis of configurable software systems.
This cost can be measured in terms of both, the number of analyzed configurations and time spent for performance measurements.
\approach helps reduce this cost in two ways---classifying changes as not performance-relevant and identifying relevant features for performance testing.
With this research question, we study how large these savings are.

\subsection{Subject Systems}
\label{sub:systems}

\newcommand{\numcoreutils}{eight\xspace}
\newcommand{\numrealworld}{three\xspace}

Our evaluation rests on synthetic, seeded, and real-world regression scenarios.
Synthetic scenarios allow us to control the setting and confounding factors, explore edge cases, and provide a qualitative view on the results, increasing internal validity.
Real-world scenarios allow us to evaluate \approach in a more realistic setting, increasing both ecological and external validity.
As a middle ground between synthetic and real-world, we also consider seeded regression scenarios based on \numcoreutils different \textsc{GNU Coreutils} tools.
\autoref{tab:cs} lists the subject systems we use in our evaluation along with the number of analyzed regression scenarios and configurations.
A \emph{regression scenario} is a pair of revisions $(v_{\text{old}}, v_{\text{new}})$ of a system.
Revision $v_{\text{old}}$ refers to the state of the system before a change and $v_{\text{new}}$ refers to the state after a change.
For the synthetic and seeded regression scenarios, we obtain $v_{\text{new}}$ by applying a hand-crafted patch to a revision $v_{\text{old}}$.
For the real-world scenarios, we select pairs of revisions from the system's revision history.
For our methodology, we aim at achieving high internal validity (instead of external validity)~\cite{MSA+26}:
We selected the subject systems to provide a proof of concept~(internal validity), rather than to make comprehensive statements on generalizability~(external validity).

\begin{table}
  \small
  \centering
  \caption{Overview of our subject systems}
  \label{tab:cs}
  \sisetup{table-alignment-mode=format, table-format=2, table-number-alignment=right}
  \begin{tabular}{@{}clrSSS[table-format=4]@{}}
    \toprule
    & Name & Domain & {Scenarios} & {$|F|$} & {$|C|$} \\
    \midrule
    \multirow{9}{*}{\rotatebox[origin=c]{90}{Synthetic}} & \textsc{Inter\textsubscript{struc}} & Interaction pattern & 1 &  1 & 2 \\
    & \textsc{Inter\textsubscript{DF}}    & Interaction pattern & 1 &  1 & 2 \\
    & \textsc{Inter\textsubscript{impl}}  & Interaction pattern & 1 &  1 & 2 \\
    & \textsc{Func\textsubscript{single}} & Regression behavior & 1 &  3 & 8 \\
    & \textsc{Func\textsubscript{accum}}  & Regression behavior & 3 &  3 & 8 \\
    & \textsc{Func\textsubscript{multi}}  & Regression behavior & 3 &  3 & 8 \\
    & \textsc{Deg\textsubscript{low}}     & Configuration space & 1 & 10 & 1024 \\
    & \textsc{Deg\textsubscript{high}}    & Configuration space & 1 & 10 & 1024 \\
    & \textsc{Deg\textsubscript{complex}} & Configuration space & 1 & 10 & 320 \\
    \midrule
    \multirow{9}{*}{\rotatebox[origin=c]{90}{Seeded}} & \textsc{basenc}  & GNU Coreutils  & 5 & 10 &   84 \\
    & \textsc{cksum}   & GNU Coreutils  & 5 & 12 &   36 \\
    & \textsc{dd}      & GNU Coreutils  & 5 & 18 & 1404 \\
    & \textsc{od}      & GNU Coreutils  & 5 & 16 &  480 \\
    & \textsc{pr}      & GNU Coreutils  & 5 & 12 & 1920 \\
    & \textsc{sort}    & GNU Coreutils  & 5 &  8 &  128 \\
    & \textsc{uniq}    & GNU Coreutils  & 5 & 14 &  384 \\
    & \textsc{wc}      & GNU Coreutils  & 5 &  5 &   32 \\
    \midrule
    \multirow{3}{*}{\rotatebox[origin=c]{90}{RW}} & \textsc{bzip2}   & Compression  & 19 &  9 & 576 \\
    & \textsc{grep}    & Unix tool    &  7 &  10 & 288 \\
    & \textsc{picoSAT} & SAT-solver   & 12 &   5 & 32  \\
    \bottomrule
  \end{tabular}
\end{table}

\paragraph{Synthetic Regression Scenarios}

We use synthetic scenarios to systematically and qualitatively assess and discuss the capabilities of \approach.
Specifically, we consider different mechanisms of how features can interact with performance-relevant code~(\textsc{Inter}), various patterns of how changes in performance behavior contribute to a regression~(\textsc{Func}), and how the structure of a configuration space and constraints between features affects \approach(\textsc{Deg}).
The code for all synthetic scenarios is contained in our replication package.

\begin{figure}
  \begin{subfigure}{0.32\textwidth}
    \includegraphics{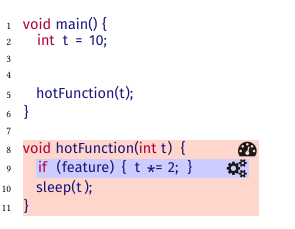}
    \caption{\textsc{Inter\textsubscript{struc}}}
  \end{subfigure}
  \begin{subfigure}{0.32\textwidth}
    \includegraphics{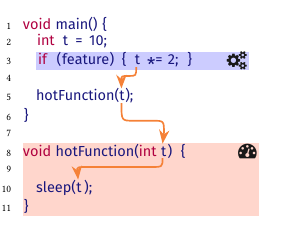}
    \caption{\textsc{Inter\textsubscript{DF}}}
  \end{subfigure}
  \begin{subfigure}{0.32\textwidth}
    \includegraphics{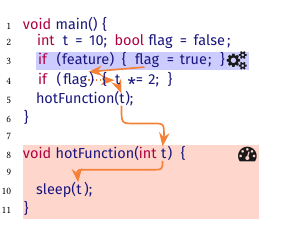}
    \caption{\textsc{Inter\textsubscript{Impl}}}
  \end{subfigure}
  \caption{
    Simplified code examples of the \textsc{Inter} scenarios demonstrating different interaction patterns between features and performance-relevant code.
    Function \code{hotFunction} is performance-relevant (\faTachometer) and code controlled by variable \code{feature} is feature code (\faGears).
    Solid lines represent data flow, the dotted line represents implicit flow.
  }
  \label{fig:synth_inter}
\end{figure}

The \textsc{Inter} scenarios cover three common mechanisms of how features can interact with performance-relevant code: structural interactions~(\textsc{Inter\textsubscript{struc}}), data-flow interactions~(\textsc{Inter\textsubscript{DF}}), and interactions via implicit data-flow~(\textsc{Inter\textsubscript{impl}}).
With these scenarios, we study the capabilities of \approach to detect which features are releated to a performance regression.
\autoref{fig:synth_inter} shows simplified code examples of these scenarios:
The regression scenario we investigate changes the initial value of variable~\code{t} that influences how long \code{hotFunction} takes to execute.
The only difference between the individual scenarios is \emph{how} feature code interacts with performance-relevant code:
In \textsc{Inter\textsubscript{struc}}, feature code interacts \emph{structurally} with performance-relevant code, i.e., the feature is part of the performance-relevant code.
In \textsc{Inter\textsubscript{DF}}, feature code modifies the value of variable~\code{t}, which in turn is passed as a parameter to \code{hotFunction}.
That is, the feature interacts with the performance-relevant code via \emph{data flow}.
In \textsc{Inter\textsubscript{impl}}, feature code influences a control-flow decision (via variable \code{flag}) that once again impacts performance-relevant code because it modifies variable~\code{t}.
Such data flow that is caused by a control-flow decision is called \emph{implicit flow}~\cite{KHH+08} and requires special handling in a data-flow analysis.
Each of these scenarios requires different capabilities to identify the relevant feature.
While \textsc{Inter\textsubscript{struc}} is easy to detect due to the structural overlap, \textsc{Inter\textsubscript{DF}} and \textsc{Inter\textsubscript{impl}} can only be detected by considering data flow---highlighting the benefit of \approach over simpler strategies.

The \textsc{Func} scenarios capture different patterns of how individual changes in performance behavior contribute towards a regression.
Each of these scenarios consists of three performance-relevant functions.
In \textsc{Func\textsubscript{single}}, a change impacting a single function with performance-relevant code triggers a regression.
In \textsc{Func\textsubscript{accum}}, we use three regression scenarios that successively impact more functions (first one, then two, and finally all three functions).
Each impacted function exhibits a change in its performance behavior, but falls below the selected threshold to trigger a regression on its own.
Only if all three are affected at the same time, the individual changes accumulate and the regression becomes noticeable.
This scenario is interesting because static analysis has no information about the magnitude of the performance regression and, thus, should report each individual interaction as a potential regression.
In \textsc{Func\textsubscript{multi}} the regression manifests only if two specific functions are impacted at the same time.
There is no change in performance behavior at all if only individual functions are affected.
This scenario poses another challenge to any static analysis tool because the exact conditions under which this regression manifests are not visible considering only data flow~(see~\autoref{sub:approach_overview}).
Consequently, we expect that \approach will report potential regressions here more often than necessary.

The \textsc{Deg} scenarios target how the structure of a configuration space and constraints between features affect \approach.
\textsc{Deg\textsubscript{low}} and \textsc{Deg\textsubscript{high}} share the same configuration space, meaning that they have the same features and configurations.
The difference is the number of configurations for which the introduced regression can be observed.
Specifically, the configuration space consist of ten optional and independent features~(i.e., $1024$ different configurations).
\textsc{Deg\textsubscript{low}} requires only very few features to be selected simultaneously to trigger the regression.
That is, the feature interaction necessary to reveal the regression has a \emph{low interaction degree}.
Such a regression is easy to detect as it can be observed in many configurations.
\textsc{Deg\textsubscript{high}}, however, requires all ten features to be selected simultaneously in order to trigger the regression.
That is, the regression requires a feature interaction with a \emph{high interaction degree} and can only be observed in a single configuration, making it very hard to detect.
The configuration space of \textsc{Deg\textsubscript{complex}} contains more complex constraints among features.
Such a constrained configuration space is more realistic, because many constraints between features exist in real-world configurable software systems~\cite{BRN+13}.

\paragraph{Seeded Regression Scenarios in Real-World Systems}

To evaluate \approach in a more realistic setting, we consider seeded regression scenarios based on \numcoreutils different tools from \textsc{GNU Coreutils}.
We chose to use tools from \textsc{GNU Coreutils} as a basis for the seeded regression scenarios because they provide a good compromise between complexity and size.
Their size allows us to understand these systems enough to seed additional regression scenarios and so that we are able to qualitatively assess the study results.
At the same time, they are complex enough to provide a realistic setting for our experiments.

For each of the \numcoreutils tools, we consider a set of five seeded regression scenarios that we have created for this purpose.
Each scenario consists of two patches: one patch simulates a performance-relevant change, and the other patch simulates the actual regression.
That way, we ensure that the seeded regression has a measurable impact on the performance behavior of the system.

\paragraph{Real-World Regression Scenarios}

For the real-world scenarios, we select \numrealworld open-source projects from different domains.
We limit ourselves to three systems due to the high cost of our experiments measuring performance of the full configuration space for multiple revisions of each system.
We focus on internal validity, and this set of three systems is sufficient and provides the necessary level of control and detail.
For each system, we consider several real-world regression scenarios.
Therefore, we select pairs of revisions that contain significant changes.
Since many of our systems are quite mature and mostly experience very small maintenance changes, these revisions are not necessarily consecutive.
For \textsc{picoSAT}, we selected revisions tagged as releases.
For \textsc{bzip2} and \textsc{grep}, we manually examined the revision history and selected commits based on their diff.
Since building a ground truth requires analyzing the full configuration space for all selected revisions of each subject system, we limit the feature model of each system to a subset of features.
That way, we reduce the costs for our experiments to a feasible level while preserving their meaningfulness, as the exact number of considered features and configurations is not the focus in this study.

\subsection{Operationalization}
\label{sub:operationalization}

\begin{figure*}
  \centering
  \hrule
  \includegraphics{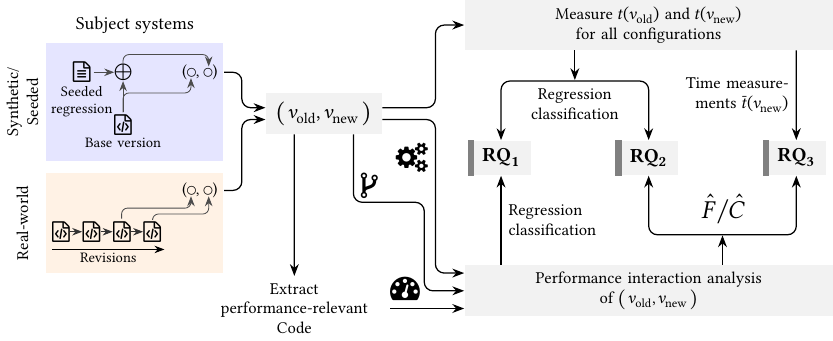}
  \hrule
  \caption{%
    Pairs of consecutive revisions of each subject system constitute the set of regression scenarios.
    For each scenario, we establish a ground truth by performing performance measurements for all configurations.
    Aditionally, we obtain a regression classification and the related features/configurations from our performance interaction analysis.
    We answer our research questions by comparing these results to the ground truth.
  }
  \label{fig:study_overview}
\end{figure*}

\autoref{fig:study_overview} provides an overview of our experiments.
For each subject system---synthetic and real-world---we consider a set of regression scenarios, that is, pairs of successive revisions $(v_{\text{old}}, v_{\text{new}})$.
In a regression scenario, revision $v_{\text{old}}$ is the current state of the system, for which we already have knowledge about its performance behavior and performance-relevant code regions.
Revision $v_{\text{new}}$ contains the code change that we want to examine for potential performance regressions.

To establish a proper ground truth for comparison with \approach, we measure the execution times for each configuration of $v_{\text{old}}$ and $v_{\text{new}}$ with $10$ repetitions to account for measurement noise.
Given a program revision $v$ and configuration $c$, we denote the set of measurements (i.e., the $10$ repetitions) of $v^c$ with $t(v^c)$ and their mean as $\bar{t}(v^c)$.
Based on these measurements, we classify a performance regression as follows:

\begin{definition}
  \label{def:regression}
  Given a threshold $\varepsilon$ and a significance level $k$, a pair of revisions $(v_{\text{old}}, v_{\text{new}})$ gives rise to a performance regression iff:

  \[ \exists\ c \in C.\:\: \abs*{\,\bar{t}(v_{\text{new}}^c) - \bar{t}(v_{\text{old}}^c)\,} \geq \func{max}\Big(\varepsilon \cdot \bar{t}(v_{\text{old}}^c), k \cdot \sigma'\Big) \]

  with $\sigma' = \func{max}\Big(\sigma\big(t(v_{\text{old}}^c)\big), \sigma\big(t(v_{\text{new}}^c)\big)\Big)$, denoting the maximum of the standard deviations of the measurements.
\end{definition}

The threshold parameter $\varepsilon$ controls the sensitivity of the regression detection, and $k$ controls the significance level.
For example, setting $\varepsilon = 0.1$ and $k=3$ means that a configuration of $v_{\text{new}}$ needs to be, at least, $10\%$ slower or faster than $v_{\text{old}}$ to be considered regressing, and the probability that such a regression is just measurement noise is $<0.3\%$.\footnote{We assume that the measurement noise is normally distributed and apply the empirical rule.}
For our experiments, we use $\varepsilon = 0.1$ and $k = 3$.
These values give us high confidence that we observe real regressions and not just measurement noise.

Our performance interaction analysis requires information about changes (\faCodeFork), features (\faGears), and performance-relevant code (\faTachometer) to find performance-relevant interactions~(see~\autoref{sec:approach}).
The change information is already encoded in the diff between $v_{\text{old}}$ and $v_{\text{new}}$.
To provide feature information to the analysis, we follow the approach outlined in \autoref{sec:background}.
As for performance-relevant code, we focus on hot code (see \autoref{sec:approach}).
To obtain hot code regions, we use two different approaches:
(1) For the synthetic scenarios, we manually annotate hot code regions.
This is possible because we specifically designed them so that we know the locations of hot code.
(2) For the \textsc{GNU Coreutils} tools and the real-world systems, we extract hot code regions through performance profiles.
Specifically, we create sample-based performance profiles\footnote{We use Linux \textsc{perf} as a sampling profiler and sample at a rate of $997~\text{Hz}$ and $10$ repetitions.} for all configurations of $v_{\text{old}}$ and extract all functions that were attributed with, at least, $5\%$ of the samples.
We use this percentage as a proxy for execution time spent inside the function's code itself (i.e., self-time).

With this setup, we address our three research questions as follows.

\paragraph{Operationalization of \ref{rq:1}}
To answer~\ref{rq:1}, we use our performance interaction analysis as a classifier for potential performance regressions and compare its results to the baseline obtained from the performance measurements (see~\autoref{def:regression}).
As \approach, by construction, can detect only regressions involving hot code, we consider only regressions that can be attributed to hot code~(i.e., if the regression is also classified as such when considering only hot code).
If we find performance-related interactions, we classify the change as a potential regression.
We compare this classification to the ground truth using a precision--recall analysis.
Since our goal is to reduce the costs of running performance measurements, our focus lies primarily on recall.
We aim for a high recall because we consider missed regressions worse than false positives.
Note that the synthetic and seeded regression scenarios do not contain any false positives, as we designed them so that all changes are performance-relevant.
For the real-world scenarios, we do not expect a high precision since any data flow between a change and hot code will result in a performance-relevant interaction, even if such a data flow does not measurably influence the performance behavior.
However, we do not consider this a problem because the information provided by \approach can still be useful for limiting the number of configurations that have to be analyzed in these cases~(see \autoref{rq:3}).

\paragraph{Operationalization of \ref{rq:2}}
The information provided by \approach must be sufficient for identifying performance regressions.
To verify this, applying \autoref{def:regression} to $\hat{C}$ instead of $C$ should yield the same classification.
For this purpose, we compare the classification of the configurations in $\hat{C}$ to the ground truth using an additional precision--recall analysis.
Note that we consider only regression scenarios that were classified as a regression by \approach, because $\hat{C}$ does not exist for the other cases.
Since the configurations in $\hat{C}$ provide less information ($\hat{C} \subseteq C$), there can never occur any false-positives~(i.e., regression scenarios that are classified as non-regressing using $C$ will never be classified as regressing using $\hat{C}$).
As a consequence, the precision will always be $1$.
Instead, we are interested in the recall: a recall $< 1$ indicates that $\hat{C}$ did not include configurations necessary to detect a regression.

\paragraph{Operationalization of \ref{rq:3}}
First, we are interested in the potential savings in terms of analyzed configurations, that is, how much smaller $\hat{C}$ is compared to $C$.
Since for this experiment $\hat{C}$ has to exist as well, we again consider only experiment runs that are classified by \approach as potential regressions.
We report the relative size $|\hat{C}|/|C|$ as well as $|\hat{F}|/|F|$ to see how the number of features flagged as related to a performance-relevant change affect the size of $\hat{C}$.
Then, we report how much measurement time would have been saved in our experiments by relating the time it takes to execute all configurations $T_C = \sum_{c \in C} \bar{t}(v_{\text{new}}^c)$ to $T_{\hat{C}} = \sum_{c \in \hat{C}} \bar{t}(v_{\text{new}}^c)$---the time required for executing only configurations from $\hat{C}$.
Last, we use the speedup~$T_C/T_{\hat{C}}$ to put these savings into perspective.

%% file: tex/results.tex
\section{Study Results}
\label{sec:results}

In this section, we report our results for the synthetic and real-world systems separately so that we can account for their individual premises in our discussion.

\subsection{Synthetic Regression Scenarios}
\label{sub:results_synthetic}

\paragraph{Results for \ref{rq:1}}
Here, we evaluate how well \approach can detect performance-relevant changes in various regression scenarios.
\autoref{tab:rq1_results} summarizes the results for \ref{rq:1}.
It shows for each scenario the key metrics of the precision--recall analysis.
Overall, \approach detects all true regressions for all our synthetic systems resulting in a recall value of $1$.
This demonstrates that \approach is capable of detecting performance-relevant changes with different characteristics and independent of the size or structure of the configuration space.
The scenarios \textsc{Func\textsubscript{accum}} and \textsc{Func\textsubscript{multi}} are noteworthy.
Here, \approach classifies all three regression scenarios as potential regressions although only one of them is a true regression.
This can be explained by the fact that, whether a regression occurs in these scenarios, depends on properties not visible to a data-flow analysis as discussed in \autoref{sub:systems}.
Still, the results of \approach are an over-approximation of the set of true performance regressions meaning that no regressions are missed.

\begin{table}
  \small
  \centering
  \caption{Synthetic systems results of the precision--recall analysis for \ref{rq:1} and \ref{rq:2} (P = positives, PP = predicted positives, TP = true positives).}
  \label{tab:rq1_results}
  \sisetup{table-alignment-mode=format, table-format=1, table-number-alignment=right}
  \begin{tabular}{@{}lSSSSS[table-format=1.2]S[table-format=1.2]SSSS[table-format=1.2]@{}}
    \toprule
    & \multicolumn{6}{c}{\ref{rq:1}} & \multicolumn{4}{c}{\ref{rq:2}} \\
    \cmidrule(lr){2-7} \cmidrule(lr{0em}){8-11}
    Project & {Scenarios} & {P} & {PP} & {TP} & {Precision} & {Recall} & {Scenarios} & {P} & {PP} & {Recall} \\
    \midrule
    \textsc{Inter\textsubscript{struc}} & 1 & 1 & 1 & 1 & 1.00 & 1.00 & 1 & 1 & 1 & 1.00 \\
    \textsc{Inter\textsubscript{DF}}    & 1 & 1 & 1 & 1 & 1.00 & 1.00 & 1 & 1 & 1 & 1.00 \\
    \textsc{Inter\textsubscript{impl}}  & 1 & 1 & 1 & 1 & 1.00 & 1.00 & 1 & 1 & 0 & 0.00 \\
    \textsc{Func\textsubscript{single}} & 1 & 1 & 1 & 1 & 1.00 & 1.00 & 1 & 1 & 1 & 1.00 \\
    \textsc{Func\textsubscript{accum}}  & 3 & 1 & 3 & 1 & 0.33 & 1.00 & 3 & 1 & 1 & 1.00 \\
    \textsc{Func\textsubscript{multi}}  & 3 & 1 & 3 & 1 & 0.33 & 1.00 & 3 & 1 & 1 & 1.00 \\
    \textsc{Deg\textsubscript{low}}     & 1 & 1 & 1 & 1 & 1.00 & 1.00 & 1 & 1 & 1 & 1.00 \\
    \textsc{Deg\textsubscript{high}}    & 1 & 1 & 1 & 1 & 1.00 & 1.00 & 1 & 1 & 1 & 1.00 \\
    \textsc{Deg\textsubscript{complex}} & 1 & 1 & 1 & 1 & 1.00 & 1.00 & 1 & 1 & 1 & 1.00 \\
    \bottomrule
  \end{tabular}
\end{table}

\paragraph{Results for \ref{rq:2}}
Next, we are concerned with whether performance regressions can be detected using only configurations from $\hat{C}$~(which is the ultimate goal to save effort and costs).
Again, \autoref{tab:rq1_results} summarizes the results for the synthetic regression scenarios.
Since this experiment can never yield false positives (as $\hat{C} \subseteq C$, there can't be any new regressions; see~\autoref{sub:operationalization}), we do not report true positives or precision values.
Any recall value smaller than $1$ indicates a regression scenario where a true regression cannot be detected using $\hat{C}$, meaning that \approach did not provide the correct information required to identify regressing configurations.
The only system where we can observe this issue is \textsc{Inter\textsubscript{impl}}.
Here, the feature triggering the regression does so via \emph{implicit flow}~(see~\autoref{sub:systems}).
\approach fails to detect this interaction between feature and hot code because implicit flows are currently not supported by the underlying data-flow analysis.
Note that this is not a conceptual limitation of our approach but follows from the choice of the particular data-flow analysis.
However, in all other scenarios, \approach provides the necessary information to detect the regression using only configurations from $\hat{C}$, which shows that it accurately identifies relevant features.

\paragraph{Results for \ref{rq:3}}
Here, we investigate how much testing effort can potentially be saved using \approach.
The results for the synthetic systems are shown in \autoref{tab:rq3_results}.
We can see that the amount of saved performance testing effort, both, in terms of tested regressions and execution time, varies significantly between different scenarios.
This can best be observed looking at the results for the \textsc{Deg} scenarios:
On the one hand, for \textsc{Deg\textsubscript{low}} $\hat{F}$ contains only a single feature leading to a severe reduction in performance testing effort (only two out of $1024$ configurations need to be tested).
On the other hand, all features are marked relevant for \textsc{Deg\textsubscript{high}}, meaning that all $1024$ configurations need to be considered during performance testing, even though a regression occurs only for a single configuration in this specific scenario.
Lastly, for \textsc{Deg\textsubscript{complex}}, $\hat{C}$ contains only $6\%$ of all configurations despite $60\%$ of features being reported as relevant.
Here, the small size of $\hat{C}$ can be explained by the fact that the structure of the configuration space constrains which features can be selected together.
From these results, we can infer that the potential cost savings depend on the structure of the configuration space as well as which features are declared relevant.
But, as the synthetic scenarios can only show some representative examples, we investigate \ref{rq:3} in more detail for the seeded and real-world scenarios later.

\begin{table}
  \small
  \centering
  \caption{
    Synthetic regression scenarios results for \ref{rq:3}.
    Grouped rows show different regression scenarios for the same system.
    A dash (--) indicates no saved performance testing effort.
  }
  \label{tab:rq3_results}
  \sisetup{table-alignment-mode=format, table-format=2, table-number-alignment=right}
  \begin{tabular}{
    @{}lSSS[table-format=4]S[table-format=4]
    S[table-format=1.2]S[table-format=1.4]
    S[table-format=4.2]S[table-format=4.2]S[table-format=3.2]@{}
  }
    \toprule
    Project & $|F|$ & $|\hat{F}|$ & $|C|$ & $|\hat{C}|$ & {$|\hat{F}|/|F|$} & {$|\hat{C}|/|C|$} & {$T_C$ ($s$)} & {$T_{\hat{C}}$ ($s$)} & {$T_C$/$T_{\hat{C}}$} \\
    \midrule
    \textsc{Inter\textsubscript{df}}    & 1 & 1 & 2 & 2 & 1.00 & 1.00 & 14.50 & 14.50 & 1.00 \\
    \textsc{Inter\textsubscript{struc}} & 1 & 1 & 2 & 2 & 1.00 & 1.00 & 14.50 & 14.50 & 1.00 \\
    \textsc{Inter\textsubscript{impl}}  & 1 & 0 & 2 & 1 & 0.00 & 0.50 & 14.50 &  6.50 & 2.23 \\
    \addlinespace[0.5em]
    \textsc{Func\textsubscript{single}} & 3 & 1 & 8 & 2 & 0.33 & 0.25 & 18.00 &  2.50 & 7.20 \\
    \addlinespace[0.5em]
    \textsc{Func\textsubscript{accum}}  & 3 & 1 & 8 & 2 & 0.33 & 0.25 & 78.00 & 13.50 & 5.78 \\
                                        & 3 & 2 & 8 & 4 & 0.66 & 0.50 & 80.00 & 34.00 & 2.35 \\
                                        & 3 & 3 & 8 & 8 & 1.00 & 1.00 & 82.00 & 82.00 & 1.00 \\
    \addlinespace[0.5em]
    \textsc{Func\textsubscript{multi}}  & 3 & 2 & 8 & 4 & 0.66 & 0.50 &  8.00 &  2.00 & 4.00 \\
                                        & 3 & 0 & 8 & 1 & 0.00 & 0.13 &  8.00 &  0.00\tablefootnote{Execution time of the selected configuration was rounded to $0.00$ by our measurement setup. Hence, no speedup can be calculated.} & \multicolumn{1}{r}{--} \\
                                        & 3 & 2 & 8 & 4 & 0.66 & 0.50 &  9.00 &  2.50 & 3.60 \\
    \addlinespace[0.5em]
    \textsc{Deg\textsubscript{low}}     & 10 &  1 & 1024 &    2 & 0.10 & 0.0020 & 3328.00 &    6.50 & 512.00 \\
    \textsc{Deg\textsubscript{high}}    & 10 & 10 & 1024 & 1024 & 1.00 & 1.00   & 2561.50 & 2561.50 &   1.00 \\
    \textsc{Deg\textsubscript{complex}} & 10 & 6 &   320 &   20 & 0.60 & 0.063  &  800.00 &   50.00 &  16.00 \\
    \bottomrule
  \end{tabular}
\end{table}

\subsection{Seeded Regression Scenarios in Real-World Systems}
\label{sub:results_coreutils}

\paragraph{Results for \ref{rq:1}}
For the seeded regressions, the results for \ref{rq:1} and \ref{rq:2} are summarized in \autoref{tab:rq1_results_seeded}.
We can see that \approach correctly identifies all scenarios as performance-relevant changes except for two cases.
For \textsc{cksum}, \approach does not detect any performance-relevant change.
The reason for this is that all functions marked as hot code are only used via function pointers and the employed data-flow analysis cannot track data flow through function pointers with sufficient precision.
As for the missed case in \textsc{dd}, the used data-flow analysis did not report any interaction between the changed code and performance-relevant code, but we were unable to determine the exact reason for this behavior.
For all other systems, \approach correctly identifies all regressions resulting in a recall of $1$, demonstrating that it is effective in detecting performance-relevant changes also in real-world code.

\paragraph{Results for \ref{rq:2}}
Notably, the information provided by \approach was sufficient to detect all but one of the considered regressions.
The one missed scenario is actually due to the first step of \approach:
It detects a performance-relevant interaction with a hot function, but the taint analysis does not reach the relevant part of that function.
In this concrete case, there is feature code within this function that influences the performance behavior, but that part of the function is not tainted.\footnote{We reported this issue to the developers of the data-flow analysis, and they confirmed that this happens due to the analysis producing unsound results in this specific case.}
Therefore, the second step correctly omits that feature from $\hat{F}$ and, thus, $\hat{C}$ contains no regressing configuration.
In all other cases, \approach provides the necessary information to detect the regression using only configurations from $\hat{C}$, which shows that the provided information is sufficiently precise such that no regressions are missed.

\begin{table}
  \small
  \centering
  \caption{Seeded regressions results of the precision--recall analysis for \ref{rq:1} and \ref{rq:2} (P = positives, PP = predicted positives, TP = true positives).}
  \label{tab:rq1_results_seeded}
  \sisetup{table-alignment-mode=format, table-format=2, table-number-alignment=right}
  \begin{tabular}{@{}lSSSSS[table-format=1.2]S[table-format=1.2]SSSS[table-format=1.2]@{}}
    \toprule
    & \multicolumn{6}{c}{\ref{rq:1}} & \multicolumn{4}{c}{\ref{rq:2}} \\
    \cmidrule(lr){2-7} \cmidrule(lr{0em}){8-11}
    Project & {Scenarios} & {P} & {PP} & {TP} & {Precision} & {Recall} & {Scenarios} & {P} & {PP} & {Recall} \\
    \midrule
    \textsc{basenc} & 5 & 5 & 5 & 5 & 1.00 & 1.00 & 5 & 5 & 5 & 1.00 \\
    \textsc{cksum}  & 5 & 5 & 0 & 0 & \multicolumn{1}{r}{-\hphantom{0}\hphantom{0}} & 0.00 & \multicolumn{1}{r}{-} & \multicolumn{1}{r}{-} & \multicolumn{1}{r}{-} & \multicolumn{1}{r}{-\hphantom{0}} \\
    \textsc{dd}     & 5 & 5 & 4 & 4 & 1.00 & 0.80 & 4 & 4 & 4 & 1.00 \\
    \textsc{od}     & 5 & 5 & 5 & 5 & 1.00 & 1.00 & 5 & 5 & 4 & 0.80 \\
    \textsc{pr}     & 5 & 5 & 5 & 5 & 1.00 & 1.00 & 5 & 5 & 5 & 1.00 \\
    \textsc{sort}   & 5 & 5 & 5 & 5 & 1.00 & 1.00 & 5 & 5 & 5 & 1.00 \\
    \textsc{uniq}   & 5 & 5 & 5 & 5 & 1.00 & 1.00 & 5 & 5 & 5 & 1.00 \\
    \textsc{wc}     & 5 & 5 & 5 & 5 & 1.00 & 1.00 & 5 & 5 & 5 & 1.00 \\

    \bottomrule
  \end{tabular}
\end{table}

\begin{figure}
  \centering
  \includegraphics{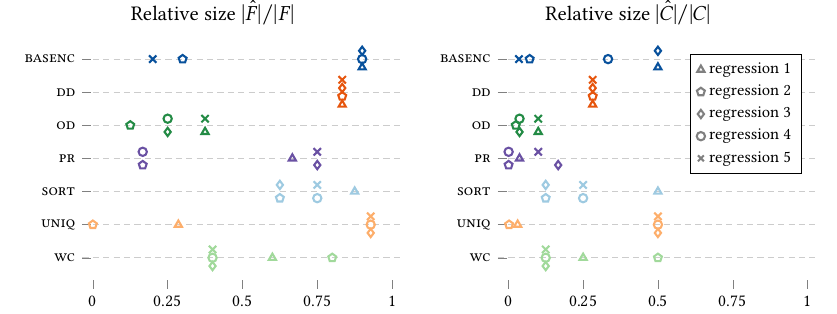}
  \vspace{-2.5ex}
  \caption{
    Seeded regressions results for \ref{rq:3}.
    For each considered coreutils tool, we show the relative sizes $|\hat{F}| / |F|$ (left) and $|\hat{C}| / |C|$ (right).
    We use different markers to enable tracking of individual regression scenarios between plots:
    the same marker in the same color corresponds to the same regression scenario.
    Note that there are only four markers for \textsc{dd} as only four regression scenarios were detected as performance-relevant~(see \autoref{tab:rq1_results_seeded}).
  }
  \label{fig:rq3_results_seeded}
\end{figure}

\begin{figure}
  \centering
  \includegraphics{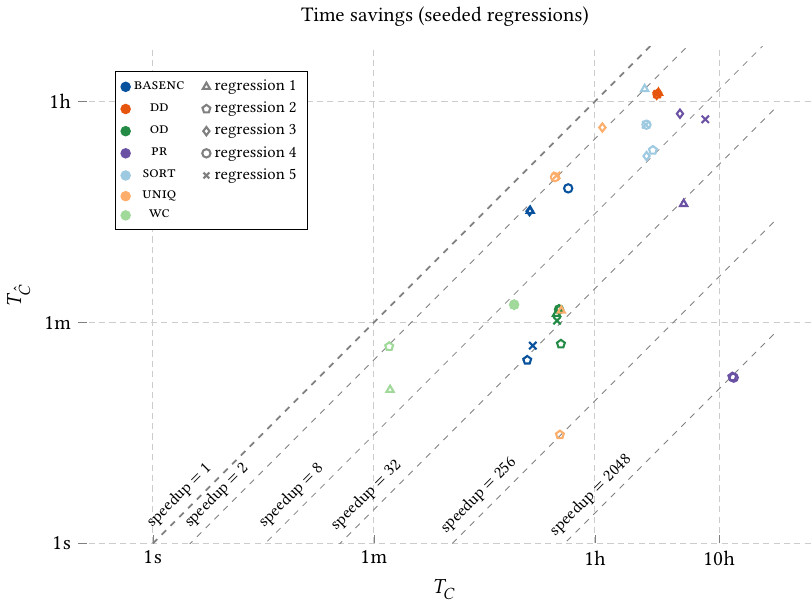}
  \caption{
    Seeded regressions execution time results for \ref{rq:3}.
    Here, we plot the time in seconds for executing all configurations ($x$-axis) to the time for executing only configurations from $\hat{C}$ ($y$-axis).
    Note that this plot uses a logarithmic scale for both axes to better accomodate the execution times of all tested systems.
    There are no values to the north-west of the diagonal $x=y$ because $\hat{C} \subseteq C$.
    Diagonal lines indicate different speedup factors $T_C/T_{\hat{C}}$.
  }
  \label{fig:rq3_time_results_seeded}
\end{figure}

\paragraph{Results for \ref{rq:3}}
We first focus on the results on cost savings in terms of the size of $\hat{C}$ depicted in \autoref{fig:rq3_results_seeded}.
In this figure, we show for each system the relative sizes of $\hat{F}$ and $\hat{C}$ compared to $F$ and $C$, respectively.
Overall, we can see that the number of features in $\hat{F}$ (and thus, the number of configurations in $\hat{C}$) varies significantly between different systems and regression scenarios.
Still, at least one feature is not considered related for all regression scenarios.
As a result, $\hat{C}$ is always at most half the size of $C$, meaning that we can always save, at least, half of the performance testing effort.
For most scenarios (22 out of 34) this reduction is even more pronounced, saving $75\%$ of configurations to test or more.
Additionally, \autoref{fig:rq3_results_seeded} visualizes the relationship between the sizes of $\hat{F}$ and $\hat{C}$:
the more features are marked as relevant, the more configurations need to be tested as the number of possible combinations of these features increases exponentially.
\autoref{fig:rq3_time_results_seeded} shows the actual execution times for executing all configurations in $C$ compared to only executing configurations from $\hat{C}$.
As the different systems have very different execution times and different numbers of configurations, the sums of execution times $T_C$ range from minutes to several hours.
To better accommodate this wide range, we use a logarithmic scale for both axes.
We can see that the speedups gained by only executing configurations from $\hat{C}$ is roughly $2$ or greater for all regression scenarios.
This makes intuitively sense, as we know from \autoref{fig:rq3_results_seeded} that we need to execute, at most, half of the configurations.
For many scenarios, the speedup is even significantly higher.
We observe the highest speedup of $\approx 2100$ for two scenarios of \textsc{pr} reducing the execution times from around $13$ hours to less than $22$ seconds.
This shows that the information provided by \approach can lead to significant cost savings in performance testing.

\subsection{Real-World Regression Scenarios}
\label{sub:results_realworld}

\paragraph{Results for \ref{rq:1}}
Finally, we present the results for the real-world regression scenarios.
\autoref{tab:rq1_results_rw} summarizes the results for \ref{rq:1}~(and \ref{rq:2}).
The first observation is that most scenarios do not contain any performance regression.
This is a case that we could not simulate well with seeded regression scenarios, but is common in real-world systems.
Therefore, it is interesting to see how well \approach can avoid false positives in these scenarios.
For \textsc{bzip2}, none of the $18$ scenarios is an actual regression.
\approach has flagged $8$ of them as potential regressions, suggesting that further performance testing should be performed in less than half of the scenarios.
For \textsc{grep} only two out of seven scenarios are actual regressions.
Here, \approach classifies four regressing scenarios as potential regressions, but detects only one of the two true regressions.
In the missed regression scenario, the introduced change consists of exactly one deleted line of code, meaning that $\crset_C^{V}$ is empty for this scenario.
Therefore, no performance-relevant interactions exist according to \autoref{def:performance_relevant_interaction}.
Only for \textsc{picosat}, most of the selected scenarios (7~out of~11) contain a regression.
Here, \approach classifies $9$ scenarios as potential regressions, including all $7$ true regressions.
These results suggest that \approach is effective in detecting performance-relevant changes also in real-world systems as long as a change doesn't solely consist of deletions.

\paragraph{Results for \ref{rq:2}}
From \autoref{tab:rq1_results_rw}, we can see that information provided by ConfFLARE was sufficient to detect regressions.
As no true regression exists for \textsc{bzip2}, the choice of $\hat{C}$ does not matter here, but it will still be interesting for \ref{rq:3} later.
For \textsc{grep}, \approach provides sufficient information to detect the one true regression.
One interesting case is regression scenario $(\texttt{ffc6e407e3}, \texttt{192e59903c})$ of \textsc{grep}, where the path reconstruction step did not terminate within a reasonable time frame.
As a consequence, no feature information is available and $\hat{C}$ contains only one configuration.
However, as this scenario is not a true regression, this does not lead to any missed regressions.
For \textsc{picosat}, all regressions can be detected using only configurations from $\hat{C}$.
From this, we can conclude that \approach provides sufficiently precise information to avoid missing regressions also in real-world systems.

\begin{table}
  \small
  \centering
  \caption{Real-world systems results of the precision--recall analysis for \ref{rq:1} and \ref{rq:2} (P = positives, PP = predicted positives, TP = true positives).}
  \label{tab:rq1_results_rw}
  \sisetup{table-alignment-mode=format, table-format=2, table-number-alignment=right}
  \begin{tabular}{@{}lSSSSS[table-format=1.2]S[table-format=1.2]SSSS[table-format=1.2]@{}}
    \toprule
    & \multicolumn{6}{c}{\ref{rq:1}} & \multicolumn{4}{c}{\ref{rq:2}} \\
    \cmidrule(lr){2-7} \cmidrule(lr{0em}){8-11}
    Project & {Scenarios} & {P} & {PP} & {TP} & {Precision} & {Recall} & {Scenarios} & {P} & {PP} & {Recall} \\
    \midrule
    \textsc{bzip2}   & 18 & 0 & 8 & 0 & 0.00 &  \multicolumn{1}{r}{--\hphantom{0}} & 8 & 0 & 0 & \multicolumn{1}{r}{--} \\
    \textsc{grep}    &  7 & 2 & 4 & 1 & 0.25 & 0.50 & 4 & 1 & 1 & 1.00 \\
    \textsc{picosat} & 11 & 7 & 9 & 7 & 0.78 & 1.00 & 9 & 7 & 7 & 1.00 \\
    \bottomrule
  \end{tabular}
\end{table}

\begin{figure}
  \centering
  \includegraphics{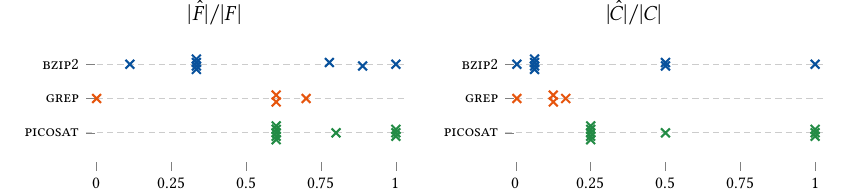}
  \vspace{-2.5ex}
  \caption{
    Real-world regressions results for \ref{rq:3}.
    For each system, we show the relative sizes $|\hat{F}| / |F|$ (left) and $|\hat{C}| / |C|$ (right).
  }
  \label{fig:rq3_results_rw}
\end{figure}

\begin{figure}
  \centering
  \includegraphics{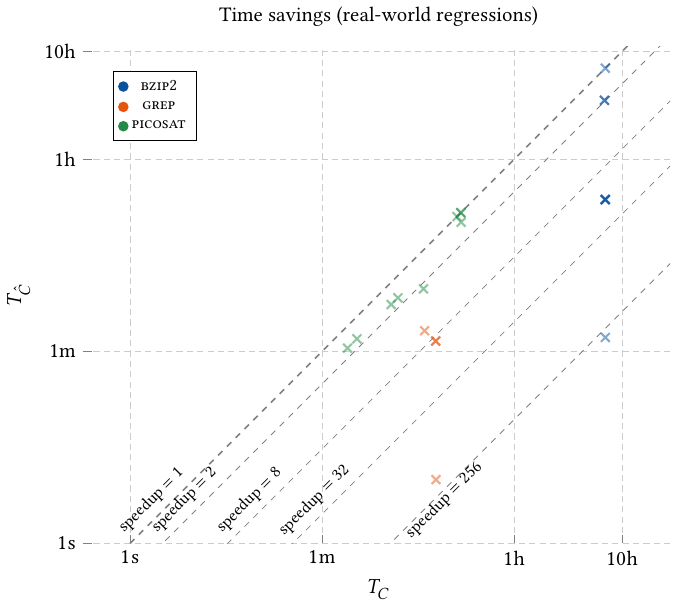}
  \caption{
    Real-world regressions execution time results for \ref{rq:3}.
    We plot the time in seconds for executing all configurations ($x$-axis) to the time for executing only configurations from $\hat{C}$ ($y$-axis).
    For a more detailed description of the plot, see \autoref{fig:rq3_time_results_seeded}.
  }
  \label{fig:rq4_time_results_seeded}
\end{figure}

\paragraph{Results for \ref{rq:3}}
For the real-world regression scenarios, we show the cost savings in terms of the size of $\hat{C}$ in \autoref{fig:rq3_results_rw}.
In contrast to the seeded regression scenarios, $\hat{F}$ contains more than half of the features in most cases.
In four cases (one for \textsc{bzip2} and three for \textsc{picosat}), all features are marked as relevant.
Consequently, $\hat{C}$ contains all configurations for these regression scenarios as well.
Still, for most scenarios, we only need to test $25\%$ of all configurations.
For the scenario of \textsc{grep} where the path reconstruction did not terminate, $\hat{F}$ is empty and, thus, $\hat{C}$ contains only a single configuration.
In \autoref{fig:rq4_time_results_seeded}, we show the execution times for the real-world systems in the same way as for the seeded regression scenarios.
For \textsc{bzip2}, the execution times are almost identical between the different regression scenarios confirming the fact that there are no regressions.
Also, the observed speedup values are in line with the relative sizes of $\hat{C}$.
A similar picture emerges for \textsc{grep}.
Only the speedup factor for the scenario where we only test one configuration is smaller than expected, suggesting that the selected configuration is slower than average.
\textsc{picosat} shows the most interesting results here:
First, almost all regression scenarios have very different execution times, supporting our claim that there exist real regressions.
The three regression scenarios where $\hat{C} = C$ unsurprisingly show no speedup at all.
However, four of the five scenarios where $\hat{C}$ contains $25\%$ of configurations, we observe a speedup of $~1.6$.
This means that we still spend $62.5\%$ of the time while executing only $25\%$ of the configurations.
This shows that the savings in terms of number of configurations do not necessarily translate directly to savings in execution time.
The same holds true for the remaining two regression scenarios of \textsc{picosat} but with different speedup factors.
These results demonstrate that, while the information provided by \approach always helps reduce the cost for performance testing, the magnitude of the reduction highly depends on the characteristics of the regression scenario in practice.

\subsection{Discussion}
\label{sub:discussion}

After presenting the results of our empirical study, we now discuss and interpret them with respect to our research questions.
Furthermore, we elaborate on how \approach advances the status quo for performance testing and how it integrates in the wider landscape of techniques for analyzing configurable software systems.

\paragraph{Identifiyng Performance Regressions and Related Features (\ref{rq:1} and \ref{rq:2})}

With our first two research questions, we intended to evaluate how well \approach can identify changes in configurable software systems that cause performance regressions and how accurately it can identify which features and configurations are impacted by these changes.
In our experiments, \approach correctly identified the majority of regressions demonstrating its ability to reliably detect performance regressions in a diverse set of circumstances.
The only case where it conceptually failed is an edge case where the change consists only of deletions, which makes it impossible to attribute the change to any code location in the analyzed version---which is one of the prerequisites for our approach to be applicable.
Other failures can be attributed to limitations of the underlying data-flow analysis, which is part of a cost-benefit trade-off that we discuss later in this section.
Furthermore, the results for the real-world regression scenarios show that \approach often correctly identifies non-regressing changes as such, which greatly helps save performance testing effort.

\rqanswer{1}{
  \approach can reliably identify changes with performance regressions in a diverse set of scenarios.
}

The main contribution of this work is to provide an automatic way of identifying which features are related to a performance-relevant change.
This information was previously only obtainable from domain experts, tedious manual work, or expensive trial-and-error testing.
To measure whether \approach provides the correct features required to observe existing performance regressions, we selected a subset of configurations to be tested based on these features.
If a regression can be detected using only the selected configurations, we consider the provided information as sufficient.
In our experiments, this was the case for all but two regression scenarios, both of which arise from limitations of the underlying data-flow analysis.
But overall, these results demonstrate that \approach is capable of accurately identifying which features are related to a performance-relevant change, which is highly valuable information for guiding performance testing.

\rqanswer{2}{
  \approach can accurately identify which features are related to a performance-relevant change in most cases.
  The accurracy and completeness of this information is subject to a cost-benefit trade-off of the underlying data-flow analysis.
}

\paragraph{Cost Saving Potential (\ref{rq:3})}

Our main motivation for this work is to make performance testing of configurable software systems more efficient by reducing the number of regression scenarios and configurations that need to be tested.
With our experiments, we demonstrated that \approach helps reduce performance testing effort in both regards:
Its first step identifies whether a regression scenario should be considered for performance testing at all.
In our experiments with real-world regression scenarios, $15$ out of $27$ non-regressing scenarios were correctly classified as such, meaning that all performance testing effort can be saved in more than half of these cases.
The information provided by \approach's second step is useful for limiting the number of configurations to be tested.
In our experiments, the fraction of configurations to be tested is on average $21\%$ for the synthetic regression scenarios and $30\%$ for the real-world regression scenarios.
Only testing configurations from $\hat{C}$ instead of $C$ resulted in speedup factors ranging from $2$ to $32$ in most cases with some outliers reaching even higher, sometimes saving hours of performance testing time.
Note that our way of constructing $\hat{C}$ from $\hat{F}$ is intentionally conservative to provide a lower bound estimate.
In practice, configurations could be selected much more intelligently, e.g., by informing sampling techniques about which features are relevant, further increasing the potential for saving measurement effort.

\rqanswer{3}{
  \approach helps significantly reduce the performance testing effort required to detect performance regressions.
  The magnitude of the savings depends on the characteristics of the regression scenario (magnitude of the regression; number of involved features) and the structure of the configuration space.
}

\paragraph{Analysis Cost-Benefit Trade-Off}

As \approach makes heavy use of data-flow analysis, the quality of its results and its cost are closely tied to the precision of the underlying analysis.
This precision enables our approach to find regressions that are hard to detect with other techniques.
However, more precise analyses are usually also more expensive to execute.
This leads to a trade-off between analysis cost and the benefit gained its results.
For example, the data-flow analysis we used in our experiments cannot track implicit flows, has limitations when tracking function pointers, and sometimes underapproximates points-to information.
Consequently, we missed some regressions in our experiments that, in theory, could have been detected with a more precise analysis, but at much higher cost.
In contrast, it is able to process real-world C/C++ code, which is a rare ability.
It is important to note that this trade-off is conceptually separate from \approach itself and, depending on the use case, different choices can be made here.

\paragraph{Integration Into Tooling Landscape}

\begin{figure}[h]
  \includegraphics{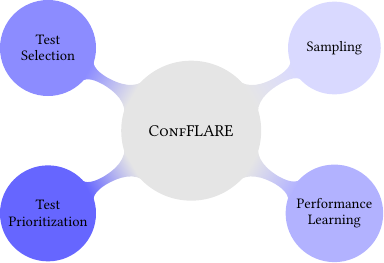}
  \caption{
    Overview of how \approach integrates with existing techniques for performance testing of configurable software systems.
  }
  \label{fig:tooling_integration}
\end{figure}

While \approach and the information it provides is useful on its own, it also complements and integrates well with existing techniques used for tackling the challenges arising from performance regression-testing of configurable software systems.
In \autoref{fig:tooling_integration}, we illustrate how \approach fits into the wider landscape of such techniques.

We already mentioned before how information about relevant features provided by \approach can be used to guide sampling or other configuration selection strategies towards configurations that are more likely to exhibit performance regressions.
For performance modeling or learning approaches, one challenge is to decide which features or combinations thereof should be included in a model.
Information about relevant features provided by \approach can help here as well as a starting point for building such models or to selectively update them after a change to the system.
The same information could also be used to select or prioritize test cases, e.g., to favor test that exercise relevant features or configurations.
The change-impact aspect of \approach complements aforementioned strategies well by skipping performance testing for entire revisions.
\approach can also complement other aspects of performance testing that are not directly related to configurations, but the execution environment.
For example, we found that one of the regression scenarios for \textsc{sort} only contains a real regression if a certain locale\footnote{The \emph{locale} specifies, amongst other things, what language and character encoding the system should use.} is set on the system.
Therefore, it is easy to miss if performance measurements are only performed on a system with the wrong locale.
By looking at the code attributed to the extracted features it was easy to spot the dependence on the system's locale.

%% file: tex/threats.tex
\subsection{Threats to Validity}
\label{sub:threats}

\paragraph{Internal Validity}

There are several factors that are important to consider for the internal validity of our study.
The most relevant one is certainly our definition of what we consider a performance regression.
The choice of the parameters $\varepsilon$ and $k$ can have an influence on the results of all research questions.
Our choice of $\varepsilon = 0.1$ and $k = 3$ ensures that only regressions with a substantial difference in execution time are considered and avoids false positives caused by measurement noise.
To further reduce the impact of measurement noise, we performed all measurements on identical hardware with an identical environment and repeated the measurements $10$ times.
Our definition of regression is also quite strict in that it flags a revision as regressing if even a single configuration regresses.
This strictness is necessary as regressions may affect only a small number of configurations.

Next is the choice of what to consider performance-relevant code.
In our study, we focus on hot code, that is, code that contributes disproportionately to the overall runtime.
We consider hot code because it can be measured automatically.
Using a sampling profiler to do so is less precise than using instrumentation measuring the exact execution times of functions.
However, we found that the overhead introduced by instrumentation is so high that sampling is the only viable choice.
For our real-world subject systems, we consider all functions that contribute more that $5$ percent to the overall execution time.
We chose that threshold because we observed a jump in function execution times around that value with most functions taking considerably less time.

\paragraph{External Validity}

As our study is the first to evaluate a novel approach for improving performance regression testing in configurable software systems, we focus on achieving high internal validity, instead of external validity~\cite{MSA+26}.
This enables us to study the \approach in a controlled setting and assess its capabilities in detail.
Further studies should then focus on how our initial results generalize~(external validity) and how \approach can be best applied in practice~(ecological validity).
Still, both, our seeded and real-world regression scenarios, use real-world software demonstrating that \approach can be applied to practical systems and not only toy examples.
Thus, we argue that our study provides a solid foundation for future research on how \approach can be utilized to its best.

%% file: tex/related_work.tex
\section{Related Work}
\label{sec:related_work}

\paragraph{Performance Evolution of Configurable Software Systems}

Considerable effort has been spent on understanding the performance behavior of configurable software systems.
Several approaches have been devised to learn models for the performance behavior of configurable software systems~\cite{CGZ23, AML+22, SCC+22, CCG+21, WAS21, VJS+21, VJS+20, HZ19, GYS+18, SGAK15}.
However, these approaches typically target individual revisions of a system.

More recently, the performance evolution of configurable software systems has gained increased attention.
In an empirical study, \citet{KMG+23} find that performance changes frequently in configurable software systems and most performance changes affect only a subset of configurations.
\citet{MAS+19} combine an iterative approach with Gaussian Process models to approximate the performance behavior of a software systems across its entire history.
They later extend this work to identify configuration-dependent performance changes across both revisions and configurations of a configurable software system~\cite{MAS+20}, although their approach works retrospectively and thus cannot be applied in a regression setting.
\citet{MAP+22} use transfer learning to reduce the number of required new training samples when transferring an existing model to new revisions.
In their study, they only evaluate the effectiveness of transfer learning across configurations and revisions using a model for the binary size of the Linux kernel and leave other questions, such as applying their approach to performance models, for future work.

Analyzing the prevalence of configuration-related performance bugs, \citet{HY+16} highlight the lack of configuration-aware regression testing suitable for performance.
They suggest that static analysis could be used to find which configurations are affected by a performance-critical change.
This is exactly the goal of \approach.

\paragraph{Static Analysis of Configurable Software Systems}

Dealing with configurability is a long-lasting challenge for static analysis.
A common technique is to address configurability in static analysis is \emph{family-based analysis}, which considers all possible configurations of a software system at once~\cite{TAK+14}.
In a large-scale study, \citet{RLJ+18} have shown that family-based analyses outperform most sampling based techniques in both efficiency and effectiveness.
Such family-based analyses can be automatically be constructed by lifting existing analyses~\cite{CTA+21}.
This can be achieved by encoding variability as choices, e.g., of types or data structures, between labeled alternatives~\cite{WKE+14}.
For example, \citet{BRT+13} present five different ways how a existing data-flow analysis can be adapted to analyze all configurations of a system.
\citet{BTR+13} use \acfi{IDE} to lift existing \acfi{IFDS} analyses to family-based analyses.
\citet{MBW14} introduce a systematic method for compositional derivation of sound family-based analyses based on abstract interpretation.
\citet{LCA+18} developed a family-based strategy for reliability analysis of configurable software systems.
More recently, \citet{DAL22} proposed a parameterized lifted analysis domain based on decision trees for analyzing numerical program families and later extended their work to deal with runtime configurable systems~\cite{DA21}.
\approach can also be classified as a family-based approach, as our focus on runtime configurable systems allows our data-flow analysis to reason about all configurations at once.

\paragraph{Regression Testing}

Regression testing has been an active research area for many years.
In this area, a lot of focus has been put on selecting~\cite{KJMG17} and prioritizing~\cite{LCZH19} test cases and minimizing test suites~\cite{KLJA18}.
Testing configurable software systems largely focuses on combinatorial interaction testing of features for which many configuration selection strategies have been proposed~\cite{MMC+14}.
Apart from that, more specialized techniques exist for selecting and prioritizing configurations~\cite{QCR08,HTM+14,HKT+16,HTL+19}, as well as generating and optimizing test suites~\cite{BLB+14,BM14}.
However, these techniques are mostly designed for single revision testing and do not take advantage of a regression testing setting.
Closest to our work is the work by \citet{QAR12}:
Similar to \approach, they use a static change impact analysis to determine which features are affected by a change.
Then, they select a set of configurations that achieves pair-wise interaction coverage for the affected features.
In contrast to our work, they only track change impact at the less precise function level~(i.e., along the call graph).
Also, they do not consider performance-relevant code, since their focus lies on traditional software (fault) testing.
Therefore, we provide a more efficient way to detect performance regressions in configurable software systems by focusing on the configurations that are affected by a change.

%% file: tex/conclusion.tex
\section{Conclusion}
\label{sec:conclusion}

Continuous performance regression testing of configurable software systems remains a challenging task due to the complex interplay between the system's configuration, its evolution, and other external factors.
In this work, we presented \approach, an approach that aims at reducing the cost for performance regression testing of configurable software systems by focusing efforts on changes and features that are likely to affect the system's performance behavior.
\approach employs a data-flow analysis to detect performance-relevant interactions between code changes and performance-relevant code and extracts feature information from interacting program paths to localize the part of the configuration space that is potentially affected by such changes.
In an empirical study using both synthetic and real-world software systems, we demonstrate that \approach is effective in detecting performance-relevant changes and in identifying relevant features for performance regression analysis.
While useful on its own, \approach can also be combined with complementary techniques, such as sampling, performance learning, or test suite optimization, to further improve the efficiency of performance regression testing in configurable software systems.
How exactly such combinations can be realized and how they perform in practice is an interesting avenue for future research.